\documentclass[lettersize,journal]{IEEEtran}
\usepackage{amsmath,amsfonts}
\usepackage{algorithmic}
\usepackage{algorithm}
\usepackage{array}
\usepackage{multirow}
\usepackage[caption=false,font=normalsize,labelfont=sf,textfont=sf]{subfig}
\usepackage{textcomp}
\usepackage{stfloats}
\usepackage{url}
\usepackage{verbatim}
\usepackage{graphicx}
\usepackage{tikz}

\usepackage{cite}
\usepackage{hyperref}
\usepackage{tcolorbox}
\tcbuselibrary{breakable} 
\usepackage{xcolor}
\usepackage{tabularx}
\usepackage{balance}
\usepackage{xcolor}
\usepackage[compact]{titlesec}

\titlespacing{\section}{0pt}{1.0ex plus 0.2ex minus 0.2ex}{0.6ex plus 0.1ex}
\titlespacing{\subsection}{0pt}{0.8ex plus 0.2ex minus 0.2ex}{0.4ex plus 0.1ex}
\titlespacing{\subsubsection}{0pt}{0.6ex plus 0.2ex minus 0.2ex}{0.3ex plus 0.1ex}

\newcommand{\legbox}[1]{%
  \textcolor[HTML]{#1}{\rule[0.3ex]{2.5mm}{2.5mm}}%
}

\begin{document}

\title{LLM-Based Digital Twin Intelligence for Application-Aware Network Selection in 6G Heterogeneous Wireless Networks}


\author{Brahim Mefgouda,~\IEEEmembership{Member,~IEEE,} Anis Bara,~\IEEEmembership{Member,~IEEE,} Lina Bariah,~\IEEEmembership{Senior Member,~IEEE,} Hang Zou,~\IEEEmembership{Member,~IEEE,} Yuzhi Yang,~\IEEEmembership{Member,~IEEE,} Mérouane Debbah,~\IEEEmembership{Fellow,~IEEE} 


\thanks{B. Mefgouda, A.Bara, L.Bariah, H.Zou, Y.Yang, and M. Debbah are with the Research Institute for Digital Future, Khalifa University, 127788 Abu Dhabi, UAE (emails: \{brahim.mefgouda,  anis.bara, lina.bariah, hang.zou, yuzhi.yang, merouane.debbah\}@ku.ac.ae). 
}

\thanks{}

}

\markboth{}%
{Mefgouda \MakeLowercase{\textit{et al.}}: LLM-Based Digital Twin Intelligence for Application-Aware Network Selection}

\maketitle

\begin{abstract}
Future 6G heterogeneous wireless networks (HWNs) are expected to support multiple radio access technologies (RATs), dynamic wireless environments, and applications with diverse quality-of-service (QoS) requirements. In such environments, network selection (NS) cannot rely only on instantaneous radio measurements or static ranking rules. Instead, access decisions must account for the evolving wireless state, service intent, packet-level QoS behavior, and candidate-RAT dynamics. This paper proposes a large language model (LLM)-based digital twin (DT) framework for stable, application-aware RAT selection under candidate-set evolution. The main idea is to shift NS from an instantaneous decision-matrix operation to a decision process over an evolving wireless DT state. The constructed DT combines site-specific geometry, Sionna RT-based propagation descriptors, ns-3 packet-level QoS emulation, service context, candidate-RAT information, and decision memory. Rather than acting as a general-purpose controller for 6G networks, the LLM is used for DT-grounded decision intelligence in this specific NS task. On top of this DT, a unified intent agent translates user and service requirements into structured decision priorities for two complementary NS branches: an LLM-assisted multi-attribute decision-making branch (MADM--LLM--NS) and a direct LLM-based ranking branch (LLM--NS). To improve decision stability, the framework further introduces history-aware adaptive normalization (HAAN) and DT-memory-driven retrieval-augmented in-context learning (RA--ICL). Numerical results show that the proposed framework reduces rank-reversal problem and unnecessary handover events, while improving service-aware QoS satisfaction compared with representative MADM-based NS baselines.
\end{abstract}


\section{Introduction}\label{sec:introduction}
\IEEEPARstart{F}{uture} 6G heterogeneous wireless networks (HWNs) are expected to support diverse radio access technologies (RATs), dense infrastructure, highly mobile user equipment (UE), and application classes with widely varying quality-of-service (QoS) requirements~\cite{saad2020vision}. In such environments, network selection (NS), which determines the target RAT during vertical handover (VHO)~\cite{wang2012mathematical}, can no longer be reduced to a simple radio-ranking problem based only on received signal strength, signal quality, or nominal throughput. Instead, an effective NS mechanism must reason over the evolving wireless state, the active service intent, and the expected service-level impact of selecting a given RAT~\cite{3gpp_ts_23_501}. However, these factors are usually dynamic, inter-dependent, and partially observable from instantaneous measurements. Consequently, traditional matrix-based NS approaches struggle to capture the wider network context and anticipate the impact of handover decisions. This motivates a shift from matrix-only NS toward digital-twin (DT)-grounded decision intelligence, where network selection relies on a cross-layer representation of the wireless system rather than over isolated instantaneous measurements. DTs provide synchronized virtual representations of the wireless environment, which integrate network mobility dynamics, traffic behavior, and service requirements within a unified view of the network. By maintaining a contextual and predictive view of the network states, they enable NS mechanisms to evaluate handover decisions conditioned on their expected QoS impact over time.

Application-aware NS becomes increasingly important in 6G environments supporting heterogeneous services with fundamentally different QoS demands. Services such as VR/AR, streaming, and background synchronization impose different trade-offs among delay, jitter, packet loss ratio (PLR), energy efficiency (EE), and access cost~\cite{zhu2021adaptive}. As a result, the most suitable RAT can vary depending on the active service, even under the same radio conditions. A RAT with strong signal quality may still fail to satisfy the latency or reliability requirements of delay-sensitive applications.

Many approaches have been proposed for NS in HWNs, including multi-attribute decision-making (MADM) \cite{roy2025fuzzy,khalili2025wcnc}, game-theoretic~\cite{van2025network,trestian2012game}, reinforcement learning \cite{allahham2022multi,9351659},  fuzzy logic schemes\cite{guo2022multiattribute,roy2025fuzzy}, and utility-based methods \cite{mefgouda2024qos}. Among them, MADM-based NS (MADM--NS) remains widely adopted because it naturally follows a decision-matrix structure, where candidate RATs are alternatives, network attributes are criteria, and user/service priorities are represented through criterion weights \cite{silva2024comprehensive}. This structure makes MADM--NS interpretable, computationally lightweight, and suitable for time-sensitive VHO decisions without requiring training data or reward design.  However, MADM--NS remains limited in dynamic and application-aware HWNs. First, it suffers from the rank reversal problem (RRP), where the ranking of candidate RATs changes with the addition or removal of alternatives, resulting in inconsistent or unstable NS decisions. Second, although criterion weights are used to represent user and service preferences, the mapping between these weights and actual QoS satisfaction is often imprecise, leading to suboptimal RAT selection across heterogeneous traffic classes. Third, because MADM--NS decisions are mainly based on instantaneous conditions, they can trigger frequent handovers (HOs) and ping-pong effects in dynamic environments. Fourth, MADM--NS typically requires manual or offline configuration of criterion weights and does not autonomously adapt to evolving user requirements or service-level objectives. 

These limitations show that improving the ranking rule alone is not sufficient. Application-aware NS requires a decision process that can operate over an evolving wireless state, account for user and service intent, and leverage past decision outcomes to improve stability under changing candidate-RAT conditions. This in turn requires a decision layer capable of interpreting service intent and adapting the selection process to the current network context.

{\color{black} Recent advances in large language models (LLMs) provide a promising direction for intent-aware NS in 6G systems~\cite{Zhou2024LLMTelecomSurvey}. Instead of acting as standalone wireless predictors, LLMs can bridge high-level service requirements and low-level network attributes by translating user and application intent into adaptive selection behavior~\cite{bariah2025waves}. This capability is useful because conventional NS approaches, including fixed MADM rules, fuzzy-logic systems, utility-based ranking, game-theoretic formulations, supervised learning models, and reinforcement-learning agents, are often tied to predefined criteria, expert rules, labeled data, or carefully designed rewards. In contrast, an LLM can adapt to new service profiles through prompting and in-context examples while preserving interpretability~\cite{Zhou2025PromptWCM}. Accordingly, the LLM is used here for intent-to-requirement translation, service-aware reasoning, and listwise ranking over DT evidence, not as a physical-layer predictor. This capability is particularly useful in DT-enabled environments, where NS depends on heterogeneous, evolving cross-layer context rather than static decision matrices alone. In the proposed framework, the LLM operates on top of the evolving wireless DT state to adapt the network-selection process to the active service requirements and candidate-RAT conditions.}

In this work, the wireless DT serves as the evolving cross-layer state used for NS and decision support. The DT supports cross-layer state representation, candidate-RAT evaluation, application-aware access decisions, and memory-based decision reuse. The focus of this paper is on decision-layer intelligence for NS, while direct physical-layer control, such as transmit-power adaptation, beam reconfiguration, scheduling, or radio resource allocation, is outside the scope of this work.
\begin{figure*} [!t]
    \centering
    \includegraphics[width=1\linewidth]{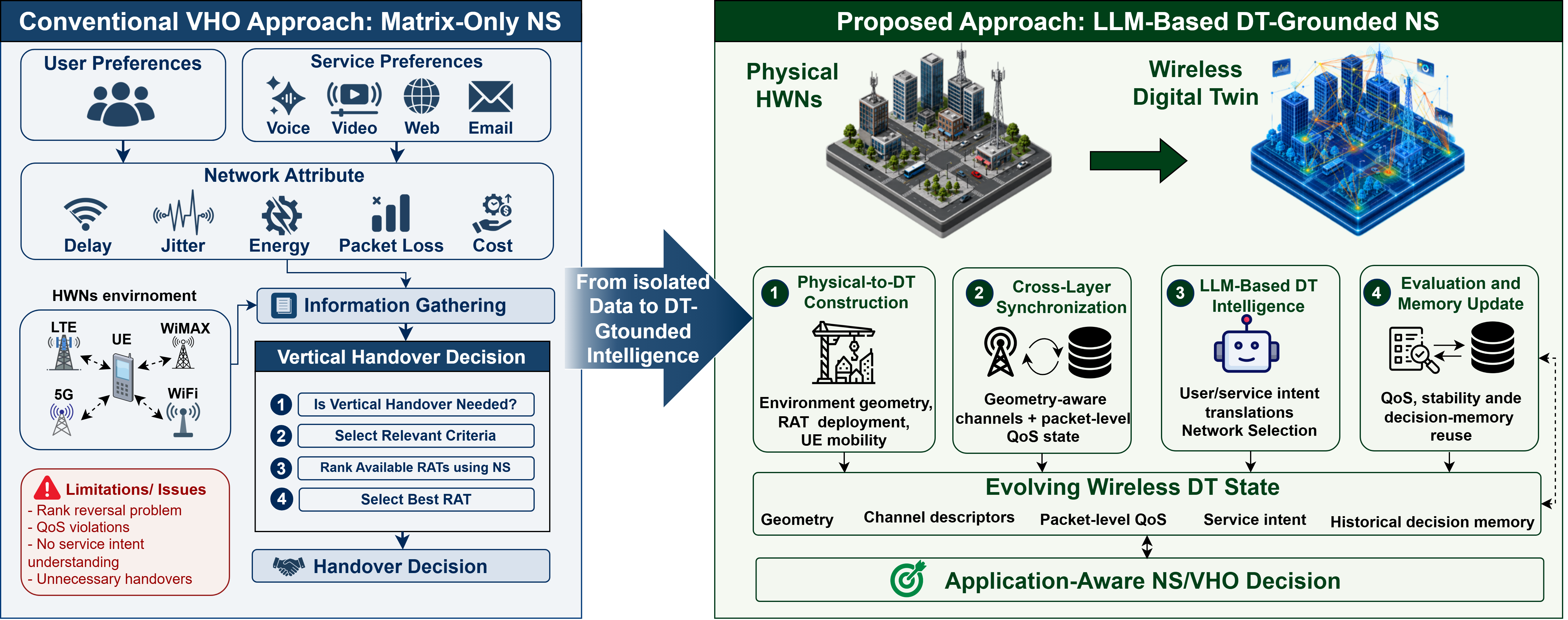}    
    \caption{Conceptual shift from matrix-level NS to DT-grounded application-aware NS over an evolving wireless DT state.}
    \label{fig:Introduction}
\end{figure*}
As shown in Fig.~\ref{fig:Introduction}, the proposed framework consists of four layers: physical-to-DT construction through site-specific 3D scene generation and Sionna RT-based propagation modeling~\cite{hoydis2023sionna}, cross-layer synchronization with ns-3 packet-level emulation~\cite{riley2010ns}, LLM-based DT intelligence, and evaluation with memory update. The main shift is from conventional matrix-level RAT ranking toward DT-grounded NS, with evolving cross-layer wireless-state-capturing geometry-aware radio descriptors, packet-level QoS, service intent, and decision memory. Unlike conventional MADM-based NS or generic LLM prompting approaches, the proposed framework combines DT-aware context, intent-driven adaptation, and memory-assisted decision reuse within a unified NS process. The main contributions are summarized as follows:

\begin{itemize}
\item  We formulate the NS problem as a DT-grounded decision task over an evolving wireless DT state. The DT maintains a cross-layer representation of the wireless environment, integrating geometry-aware channel descriptors, packet-level QoS states, service intent, and historical decision memory.
\item We design a unified intent agent (UIA) that translates user/service context into structured NS requirements. Based on this profile, we develop two DT-conditioned NS branches. In MADM--LLM--NS, the LLM derives service-aware criterion weights for deterministic MADM ranking, while in LLM--NS, the LLM performs direct listwise RAT ranking using service-specific in-context examples and DT-memory retrieval.
\item We introduce history-aware adaptive normalization (HAAN) to reduce candidate-set sensitivity and combine it with retrieval-augmented in-context learning (RA--ICL) over DT memory to stabilize LLM-based ranking under candidate-RAT evolution.

\item We provide both analytical and networking-level evaluation of the proposed framework. Specifically, we analyze candidate-set sensitivity and HAAN robustness, and evaluate the proposed framework against representative MADM--NS baselines across five 3GPP-aligned service profiles using RRP robustness, QoS satisfaction, unnecessary VHO rate, and computational complexity.
\end{itemize}

{\color{black} The novelty of this work is not a simple combination of DTs, MADM, and LLMs; it is the reformulation of NS from an instantaneous decision-matrix operation into an LLM-assisted DT-grounded decision process, where service intent, packet-level DT evidence, and decision memory are jointly used to support stable NS under candidate-set evolution. Classical MADM--NS~\cite{wang2012mathematical,obayiuwana2017network,silva2024comprehensive} ranks candidate RATs from instantaneous decision matrices, differing mainly in the weighting, normalization, or utility transformation employed. In contrast, the proposed framework derives service-aware criterion priorities from user/service intent at runtime, replacing manually configured weights. Existing DT-assisted optimization, selection, and handover studies~\cite{bilen2022proof,tang2025digital,mefgouda2025balancing,cakir2026digital} mainly use the DT for replication, prediction, emulation, or task-specific decision support. Here, the DT serves as the decision state itself, jointly capturing geometry-aware channel descriptors, packet-level QoS, service intent, candidate-RAT context, and decision memory, and this evaluated memory is reused to keep the ranking of retained candidates stable when RATs join or leave the candidate set. Likewise, LLM-for-network-management studies~\cite{Zhou2024LLMTelecomSurvey,Jiang2024LLMMultiAgentWCM,Zhou2025PromptWCM} mainly use LLMs as assistants, interfaces, or orchestration modules, whereas in this work the LLM is confined to intent-to-requirement translation and structured listwise ranking grounded in packet-level DT evidence. To the best of our knowledge, this DT-grounded formulation of candidate-set-aware RAT selection has not been explicitly addressed in prior NS studies.}

The remainder of this paper is organized as follows. Section~\ref{sec:related_work} reviews the related work, and Section~\ref{sec:system_model} presents the DT-grounded system model and problem formulation. Sections~\ref{sec:physical_to_dt}--\ref{sec:evaluation_memory_update} detail the four framework layers, from physical-to-DT construction to LLM-based DT intelligence and memory update. Section~\ref{sec:eval_setup} describes the simulation setup, Section~\ref{sec:numerical_results} reports the numerical results, and Section~\ref{sec:conclusion} concludes the paper.

\section{Related Work}
\label{sec:related_work}
This paper is related to several research directions, including wireless DTs, NS/VHO in HWNs, LLM-enabled wireless intelligence, and standard-aligned service and mobility evaluation. While these directions provide important foundations, none of them individually addresses intent-aware and memory-assisted NS over an evolving wireless DT state.
\subsection{Wireless DT and Cross-Layer Network Representation}
Wireless DTs have been introduced to represent and optimize communication systems through virtual counterparts of physical wireless environments~\cite{mihai2022digital}. Recent works have explored DT-enabled scheduling, user association, wireless optimization, site-specific ray tracing, and packet-level emulation~\cite{xu2023digital,xu2023digital2,jia2023new,zhou2023digital,hoydis2023sionna}, demonstrating the ability of DTs to capture realistic radio and protocol-level behavior. However, most existing DT studies focus on network modeling and optimization rather than application-aware NS and VHO decision making. In particular, they do not jointly incorporate service intent, candidate-RAT dynamics, and historical decision context for stable NS over an evolving DT state.
\subsection{Network Selection (NS) in HWNs}
\label{subsec:rw_ns_vho}
Several MADM--NS approaches have been proposed in literature \cite{wang2012mathematical,silva2024comprehensive}, including the technique for order preference by similarity to an ideal solution (TOPSIS), simple additive weighting (SAW), grey relational analysis (GRA), VIKOR, combined compromise solution (CoCoSo), measurement of alternatives and ranking according to compromise solution (MARCOS), and their variants~\cite{khalili2025tvt,dos2025network,mefgouda2021cocoso,ndashimye2021multi,khalili2025wcnc,mefgouda2024qos}.  In a generic MADM--NS pipeline, candidate RATs are organized into a decision matrix, normalized to handle heterogeneous criteria, weighted to reflect user/service priorities, processed by a ranking operator, and finally mapped to a target-RAT decision for association or VHO execution. Accordingly, prior works addressing MADM--NS limitations can be grouped into normalization-based and weighting-based solutions.


Normalization-based solutions modify the scaling or utility transformation applied to heterogeneous criteria, since MADM rankings can be sensitive to normalization choices~\cite{jahan2015state,tu2025analytic}. Examples include max--min and monotonic-utility reformulations of TOPSIS closeness~\cite{senouci2016topsis,senouci2016utility}, closeness-index utility  matrices~\cite{chandavarkar2016simplified}, modified VIKOR  normalization for RRP and unnecessary-VHO reduction~\cite{baghla2018vikor}, fuzzy Manhattan-distance normalization ~\cite{mansouri2019new}, and sigmoid-based transformations for MARCOS~\cite{mefgouda2024qos}. These solutions can reduce, and in some cases remove, RRP and the unnecessary VHO events caused by ranking instability. However, they do not determine which criteria should dominate for a given service and therefore cannot, by themselves, ensure QoS satisfaction across heterogeneous traffic classes or automate service-intent translation.


Weighting-based solutions target preference alignment by replacing the standard weighting techniques with metaheuristic, analytical, or hybrid weighting schemes. For instance, metaheuristic methods, including genetic algorithms~\cite{almutairi2018genetic}, particle swarm optimization~\cite{almutairi2021particle,radouche2021new}, and whale optimization~\cite{mefgouda2023new}, optimize the weight vector to improve candidate discrimination or service-oriented selection. Other schemes use uncertainty-aware or hybrid weighting, including intuitionistic normal fuzzy AHP~\cite{yu2020novel}, fuzzy analytic hierarchy process (AHP) with entropy weighting~\cite{yu2018heterogeneous}, fuzzy ANP~\cite{radouche2020network}, and fuzzy AHP variants~\cite{krishnan2025algorithm,singh2025intelligent}. These methods can improve the match between the selected RAT and predefined user or service preferences. However, they do not directly address the RRP, and their weights remain tied to optimization objectives, membership functions, expert judgments, or data statistics. Thus, these solutions improve preference alignment, but they do not provide autonomous service-intent translation at runtime.

Overall, normalization-based methods mainly improve RRP robustness, whereas weighting-based methods mainly improve QoS and preference alignment. However, most existing solutions in both directions remain tied to an instantaneous decision matrix and do not explicitly reuse historical decision context. Moreover, neither direction, in its conventional form, provides a unified mechanism that jointly captures the evolving wireless state, packet-level QoS, service intent, candidate-RAT context, and decision memory. This motivates a DT-grounded formulation in which NS is performed over an evolving cross-layer state, conditioned on service intent, and stabilized through reusable decision memory.

\subsection{LLMs for Telecommunications and Wireless Intelligence}
LLMs and foundation models have recently been investigated for telecom intent interpretation, network automation, and wireless intelligence applications~\cite{Zhou2024LLMTelecomSurvey,zou2025telecomgpt,Boateng2026LLMCommSurvey,zou2026large}. Recent studies have explored LLM-based network agents, multi-agent coordination, and prompt-conditioned wireless intelligence for adaptive 6G operation~\cite{Jiang2024LLMMultiAgentWCM,Zhou2025PromptWCM,zou2025access}. However, existing works generally treat the LLM as a standalone assistant or semantic interface rather than as a decision layer operating over an evolving wireless DT state. In particular, they do not jointly capture geometry-aware channel conditions, packet-level QoS, service intent, candidate-RAT evolution, and historical decision context for application-aware NS and VHO decision making.

\subsection{Standard-Aligned Service and Mobility Evaluation}
\label{subsec:rw_standards}
Standardization provides the service and mobility context needed to evaluate NS/VHO decisions. 3GPP QoS specifications define service classes and QoS characteristics with different sensitivities to delay, jitter, PLR, and reliability~\cite{3gpp_ts_23_501}. 3GPP performance-management specifications further define mobility-related indicators, including HO execution, service continuity, and unnecessary HO behavior~\cite{3gpp_ts_28_552}. IEEE~802.21 provides a technology-independent framework for media-independent HO across heterogeneous access technologies~\cite{taniuchi2009ieee}. However, these standards do not prescribe a concrete NS algorithm, nor do they define how service intent, packet-level QoS, geometry-aware channel descriptors, and historical decision context should be fused into a unified DT-grounded decision state.

\subsection{Positioning of This Work}
\label{subsec:rw_positioning}

The reviewed directions are complementary but incomplete when considered separately. Wireless DTs provide an evolving cross-layer representation, but existing DT studies rarely couple this representation with intent-aware multi-RAT NS. NS/VHO methods provide mature ranking, weighting, and mobility-control mechanisms, but most of them operate over instantaneous decision matrices, fixed preference models, or learned policies. LLM-enabled wireless intelligence provides semantic intent interpretation and adaptive reasoning, but existing LLM-oriented studies are rarely grounded in packet-level wireless DT states or evaluated using NS/VHO stability metrics. Finally, 3GPP and IEEE standards provide service classes, HO concepts, and mobility-evaluation references, but they do not define how DT state, service intent, and decision memory should be combined for RAT selection.

This paper addresses this gap by formulating NS as a DT-grounded decision-intelligence problem. The wireless DT provides the evolving cross-layer state, including environment geometry, geometry-aware channel descriptors, packet-level QoS, service profile, candidate-RAT context, and decision memory. The LLM-based layer translates user and service intent into structured decision behavior, while DT memory is reused to improve ranking consistency under candidate-set evolution. Thus, the proposed framework is not an unconstrained LLM ranker over simulated data; it is an intent-aware decision layer operating over an evolving wireless DT state.

\begin{figure*}[!htbp]
    \centering
    \includegraphics[width=0.85\linewidth]{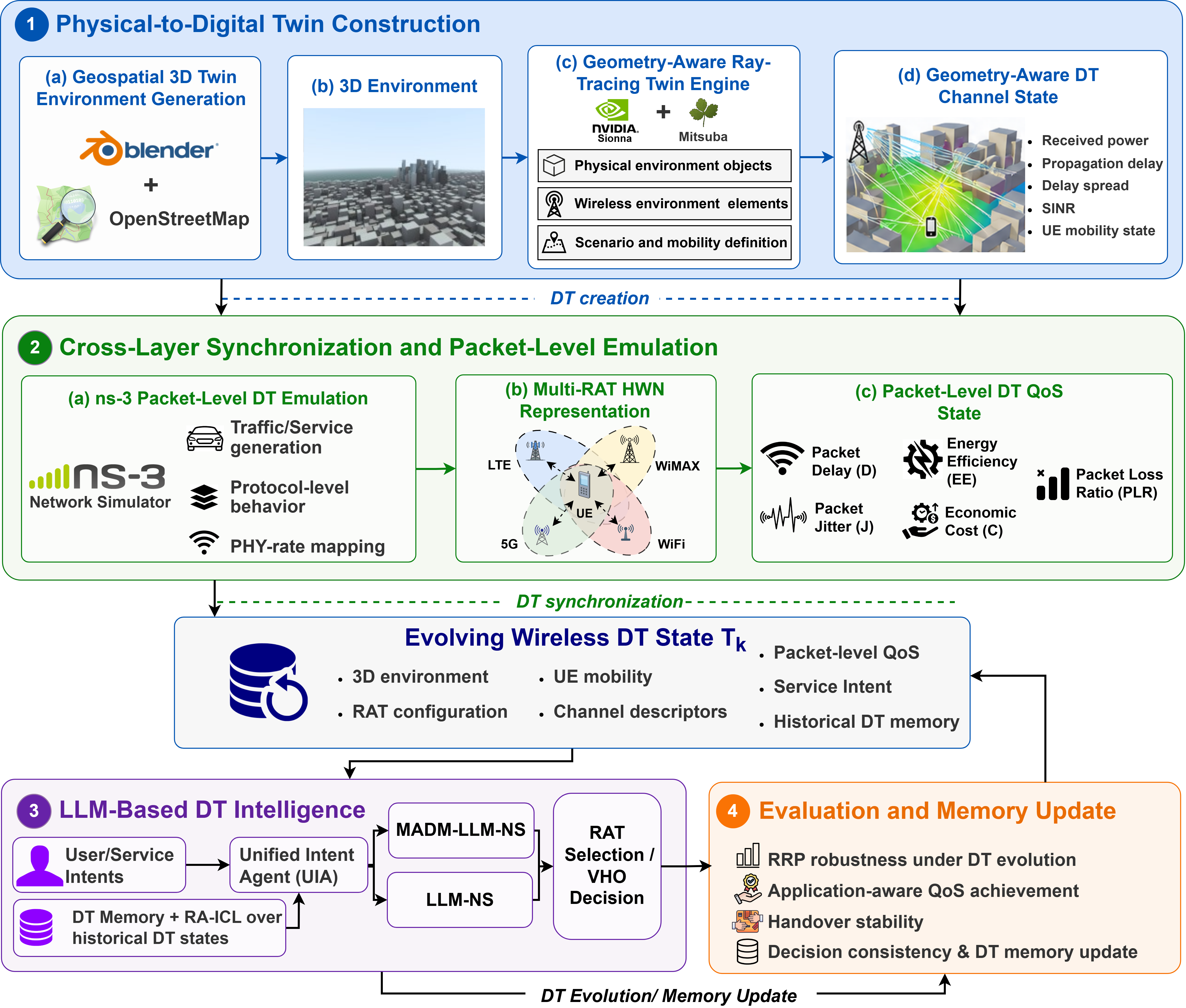}
    \caption{Proposed LLM-based DT-grounded framework for application-aware NS in HWNs.}

        \label{fig:proposed_framework}
\end{figure*}

{ \color{black} In summary, three properties jointly distinguish this work. First, unlike normalization- or weighting-based MADM--NS, the decision operates over an evolving DT state and reuses evaluated decision memory rather than refining a single static ranking rule. Second, unlike DT-based wireless optimization, the DT is used as the decision state for intent-aware RAT selection rather than for scheduling or resource allocation. Third, unlike LLM-for-network-management studies, the LLM is grounded in packet-level DT evidence, restricted to intent translation and listwise ranking, and evaluated through NS/VHO stability metrics. None of the reviewed directions provides this combination.}

\section{System Overview and DT-Grounded Problem Formulation}
\label{sec:system_model}

This section presents the proposed DT-grounded framework for application-aware NS. Instead of relying on an instantaneous decision matrix, each access decision is conditioned on an evolving cross-layer DT state capturing geometry, RAT deployment, UE position, channel descriptors, packet-level QoS, service intent, and decision memory. The DT therefore supports application-aware RAT selection and memory-assisted decision reuse.

Fig.~\ref{fig:proposed_framework} illustrates the proposed four-layer framework. The first layer constructs the site-specific physical-to-DT representation and extracts geometry-aware channel descriptors, while the second layer maps them to service-level QoS states through ns-3 packet-level emulation. The third layer performs LLM-based intent translation and DT-conditioned NS, and the fourth layer evaluates the selected RAT and updates the DT memory.



\subsection{Heterogeneous Wireless Network Model}
\label{subsec:hwn_model}




We consider an HWN deployed over a region $\mathcal{D}\subset\mathbb{R}^{3}$. The network supports $R$ RATs, indexed by $\mathcal{R}=\{1,2,\ldots,R\}$, including LTE, 5G NR, Wi-Fi, WiMAX, and other access technologies. Each RAT $r\in\mathcal{R}$ is associated with a set of access nodes $\mathcal{B}_{r}$, where an access node represents a base station, access point, or radio site.

Each access node $b\in\mathcal{B}_{r}$ is characterized by its location $\mathbf{q}_{r,b}\in\mathbb{R}^{3}$, transmit power $P_{r,b}^{\mathrm{tx}}$, carrier frequency $f_r$, bandwidth $W_r$, antenna configuration, and noise figure $F_r$. The corresponding RAT configuration is defined as
\begin{equation}
    \boldsymbol{\eta}_{r}
    =
    \left(
    f_r,
    W_r,
    F_r,
    \left\{
    P_{r,b}^{\mathrm{tx}},
    \mathbf{q}_{r,b}
    \right\}_{b\in\mathcal{B}_{r}}
    \right).
    \label{eq:rat_config}
\end{equation}
For clarity, the formulation is presented for a single multi-RAT UE. The extension to multiple UEs is straightforward by introducing a UE index into the corresponding states and decision variables.

Time is divided into decision epochs \(k\in\{1,\ldots,K\}\), each associated with a UE position \(\mathbf{p}_{k}\). As the UE moves, the changing positions lead to variations in channel conditions, QoS states, and the candidate RAT set \(\mathcal{A}(k)\).

\begingroup
\color{black}
At each decision epoch, a RAT is considered a feasible candidate only if at least one of its access nodes satisfies coverage, propagation, and access-admission constraints. Let
\(d_{k,b}^{(r)}=\|\mathbf{p}_k-\mathbf{q}_{r,b}\|\) denote the distance between the UE and access node \(b\) of RAT \(r\). The received power from node \(b\) is written in linear scale as
\begin{equation}
P_{k,\mathrm{rx}}^{(r,b)}
=
P_{r,b}^{\mathrm{tx}} G_{k,b}^{(r)},
\label{eq:rx_power_candidate}
\end{equation}
where \(G_{k,b}^{(r)}\) is the ray-traced channel gain, including the propagation and antenna effects captured by the DT. The corresponding SINR is
\begin{equation}
\gamma_{k}^{(r,b)}
=
\frac{P_{k,\mathrm{rx}}^{(r,b)}}
{N_0 W_r F_r + I_k^{(r,b)}} .
\label{eq:sinr_candidate}
\end{equation}

The feasibility indicator of access node \(b\) for RAT \(r\) is defined as
\begin{equation}
\alpha_{k,b}^{(r)}
=
\mathbb{I}
\left\{
\begin{array}{l}
d_{k,b}^{(r)} \leq R_{r,b}^{\mathrm{cov}},\\
L_{k,b}^{(r)} \geq 1,\\
P_{k,\mathrm{rx}}^{(r,b)} \geq P_{r,\min}^{\mathrm{rx}},\\
\gamma_{k}^{(r,b)} \geq \gamma_{r,\min},\\
\mu_{k,b}^{(r)} = 1
\end{array}
\right\},
\label{eq:node_feasibility}
\end{equation}
where \(R_{r,b}^{\mathrm{cov}}\) is the coverage radius, \(L_{k,b}^{(r)}\) is the number of valid ray-traced paths, \(P_{r,\min}^{\mathrm{rx}}\) is the minimum received-power threshold, \(\gamma_{r,\min}\) is the minimum SINR threshold, and \(\mu_{k,b}^{(r)}\in\{0,1\}\) is an access-admission flag capturing UE capability, operator policy, subscription, and admission availability.

Accordingly, the feasible access-node set of RAT \(r\) is
\begin{equation}
\mathcal{B}_{r}^{\mathrm{feas}}(k)
=
\left\{
b\in\mathcal{B}_r:
\alpha_{k,b}^{(r)}=1
\right\},
\label{eq:feasible_bs_set}
\end{equation}
and the candidate RAT set is explicitly given by
\begin{equation}
\mathcal{A}(k)
=
\left\{
r\in\mathcal{R}:
\mathcal{B}_{r}^{\mathrm{feas}}(k)\neq\emptyset
\right\}.
\label{eq:candidate_rat_set}
\end{equation}
For each feasible RAT, the serving access node is selected as the one providing the strongest received power,
\begin{equation}
b_k^{(r)}
=
\arg\max_{b\in\mathcal{B}_{r}^{\mathrm{feas}}(k)}
P_{k,\mathrm{rx}}^{(r,b)} .
\label{eq:serving_node_selection}
\end{equation}
The RAT-level channel descriptor \(\mathbf{h}_{k}^{(r)}\) used in the DT state is then obtained from the selected serving node \(b_k^{(r)}\). Thus, \(\mathcal{A}(k)\) is not an arbitrary subset of RATs; it is the set of RATs that have at least one admissible access node satisfying the coverage, ray-tracing link-feasibility, received-power, SINR, and access-admission constraints in~\eqref{eq:node_feasibility}--\eqref{eq:candidate_rat_set}.
\endgroup



A VHO event occurs when the selected serving RAT changes between two consecutive decision epochs,
\begin{equation}
    H_k^{\mathrm{VHO}}
    =
    \mathbb{I}
    \left\{
    a_k\neq a_{k-1}
    \right\},
    \label{eq:vho_indicator}
\end{equation}
where $H_k^{\mathrm{VHO}}$ is a binary VHO flag that equals one when the selected RAT changes between two consecutive epochs and zero otherwise, and $\mathbb{I}\{\cdot\}$ denotes the indicator function.


\subsection{Cross-Layer DT State Representation}
\label{subsec:dt_state_representation}

The physical environment is represented by a site-specific 3D scene, $\mathcal{E} =\left( \mathcal{O},\mathcal{G},\mathcal{M}_{\mathrm{mat}}\right)$, where $\mathcal{O}$ is the set of physical objects, $\mathcal{G}$ denotes their geometric layout, and $\mathcal{M}_{\mathrm{mat}}$ contains the associated radio-material properties.

For each RAT $r$ and UE position $\mathbf{p}_k$, the geometry-aware propagation component of the DT produces a compact channel descriptor, as follows:
\begin{equation}
    \mathbf{h}_{k}^{(r)}
    =
    \mathcal{F}_{\mathrm{RT}}
    \left(
    \mathcal{E},
    \boldsymbol{\eta}_{r},
    \mathbf{p}_{k}
    \right),
    \label{eq:rt_mapping}
\end{equation}
where $\mathcal{F}_{\mathrm{RT}}(\cdot)$ denotes the site-specific propagation mapping. In the implementation, this mapping is realized using Sionna RT. The descriptor is written as,
\begin{equation}
    \mathbf{h}_{k}^{(r)}
    =
    \left[
    G_{k}^{(r)},
    \tau_{k}^{(r)},
    \sigma_{\tau,k}^{(r)},
    L_{k}^{(r)}
    \right]^{\top},
    \label{eq:channel_descriptor}
\end{equation}
where $G_{k}^{(r)}$ is the channel gain, $\tau_{k}^{(r)}$ is the first-arrival delay, $\sigma_{\tau,k}^{(r)}$ is the RMS delay spread, and $L_{k}^{(r)}$ is the number of valid propagation paths. Received power and SINR are treated as derived link indicators computed during Sionna RT-to-ns-3 synchronization from $\mathbf{h}_{k}^{(r)}$, RAT configuration, and interference assumptions.
The geometry-aware DT channel state is
\begin{equation}
    \mathbf{H}_{k}
    =
    \left\{
    \mathbf{h}_{k}^{(r)}:
    r\in\mathcal{A}(k)
    \right\}.
    \label{eq:dt_channel_state}
\end{equation}

\subsection{Packet-Level QoS State and Decision Matrix}
\label{subsec:qos_derivation}

The geometry-aware channel descriptors are mapped to packet-level QoS
indicators through ns-3 emulation,
\begin{equation}
    \mathbf{x}_{k}^{(r)}
    =
    \mathcal{F}_{\mathrm{net}}
    \left(
    \mathbf{h}_{k}^{(r)},
    \boldsymbol{\eta}_{r},
    \mathbf{S}(k)
    \right),
    \label{eq:packet_qos_mapping}
\end{equation}
where $\mathcal{F}_{\mathrm{net}}$  denotes the packet-level network emulation operator realized via ns-3. The resulting QoS vector is
\begin{equation}
\mathbf{x}_{k}^{(r)}
=
\left[
D_k^{(r)},
J_k^{(r)},
\mathrm{PLR}_k^{(r)},
\mathrm{EE}_k^{(r)},
C_k^{(r)}
\right]^{\top},
\label{eq:qos_vector}
\end{equation}
where $D_k^{(r)}$, $J_k^{(r)}$, $\mathrm{PLR}_k^{(r)}$, $\mathrm{EE}_k^{(r)}$, and $C_k^{(r)}$ denote delay, jitter, packet-loss ratio, EE, and access cost, respectively. Delay, jitter, PLR, and access cost are treated as cost criteria, while EE is a benefit criterion. The packet-level DT QoS state is represented as \vspace{-5pt}
\begin{equation}
    \mathbf{Q}_{k}
    =
    \left\{
    \mathbf{x}_{k}^{(r)}:
    r\in\mathcal{A}(k)
    \right\}.
    \label{eq:qos_state}
\end{equation}
At epoch $k$, the feasible RATs form the DT-derived decision matrix as

\begin{equation}
    \mathbf{DM}(k)
    =
    \begin{bmatrix}
    \mathbf{x}_{k}^{(r_1)} &
    \cdots &
    \mathbf{x}_{k}^{(r_{N_k})}
    \end{bmatrix}^{\top},
    r_i\in\mathcal{A}(k),
    N_k=|\mathcal{A}(k)|.
    \label{eq:decision_matrix}
\end{equation}
In conventional MADM--NS, $\mathbf{DM}(k)$ is typically the complete decision input. In the proposed formulation, it is only a packet-level projection of the richer DT state.



\subsection{Unified Intent Profile and DT Decision Context}
\label{subsec:uia_decision_context}

The UIA is the shared intent-to-requirement interface of the proposed framework. The input consists of the user/operator preference context $\mathbf{U}(k)$, the active service profile $\mathbf{S}(k)$, and the criterion set $\mathcal{C}$. When context-adaptive weighting is enabled, the UIA can also receive a compact summary of the current candidate landscape extracted from $\mathbf{DM}(k)$, such as per-criterion ranges or averages. The UIA does not select a RAT, its output is a structured requirement profile that defines how the downstream NS module should interpret and prioritize the criteria. The UIA mapping is written as
\begin{equation}
    \mathbf{q}(k)
    =
    \mathcal{L}_{\mathrm{UIA}}
    \left(
    \mathbf{U}(k),
    \mathbf{S}(k),
    \mathcal{C},
    \boldsymbol{\zeta}(k)
    \right),
    \label{eq:req_profile}
\end{equation}
where $\mathcal{L}_{\mathrm{UIA}}(\cdot)$ denotes the LLM-based UIA and $\boldsymbol{\zeta}(k)$ denotes the optional compact summary of $\mathbf{DM}(k)$. If no adaptive summary is used, $\boldsymbol{\zeta}(k)$ is omitted. The unified requirement profile is expressed as
\begin{equation}
    \mathbf{q}(k)
    =
    \left\{
    \left(
    t_j(k),
    w_j(k),
    \ell_j(k),
    \chi_j(k)
    \right)
    \right\}_{j=1}^{|\mathcal{C}|},
    \label{eq:uia_output}
\end{equation}
where $t_j(k)\in\{\mathrm{benefit},\mathrm{cost}\}$ is the criterion type, $w_j(k)$ is the normalized criterion weight, $\ell_j(k)$ is the qualitative importance label, and $\chi_j(k)$ is a short criterion-level justification. The QoS target or tolerance value of criterion $c_j$ is denoted by $\theta_j^{\mathbf{S}(k)}$ and is obtained from the active service-profile configuration, not inferred as a free output of the UIA. Accordingly, the weights should satisfy
\begin{equation}
    w_j(k)\geq 0,
    \qquad
    \sum\nolimits_{j=1}^{|\mathcal{C}|} w_j(k)=1.
    \label{eq:uia_weight_sum}
\end{equation}
Thus, $\mathbf{q}(k)$ is the machine-readable form of the service/user intent. It replaces manually fixed weights in the MADM branch and conditions the direct LLM ranker in the LLM--NS branch. At epoch $k$, the evolving wireless DT state is defined as
\begin{equation}
    \mathcal{T}_{k}
    =
    \left(
    \mathcal{E},
    \mathcal{R},
    \mathbf{p}_{k},
    \mathcal{A}(k),
    \mathbf{H}_{k},
    \mathbf{Q}_{k},
    \mathbf{U}(k),
    \mathbf{S}(k),
    \mathbf{q}(k),
    \mathcal{M}_{k}
    \right),
    \label{eq:dt_state}
\end{equation}
where $\mathcal{M}_{k}$ is the DT memory. The corresponding DT-derived decision context is
\begin{equation}
    \mathcal{Z}_{k}
    =
    \left(
    \mathbf{DM}(k),
    \mathbf{H}_{k},
    \mathbf{U}(k),
    \mathbf{S}(k),
    \mathbf{q}(k),
    \mathcal{M}_{k}
    \right).
    \label{eq:decision_context}
\end{equation}

The DT memory stores previous decision contexts and outcomes. A memory entry generated after decision epoch  $k$ is
\begin{equation}
    m_k
    =
    \left(
    \boldsymbol{\phi}_{k},
    \mathbf{U}(k),
    \mathbf{S}(k),
    \mathbf{q}(k),
    \mathcal{A}(k),
    \mathbf{H}_{k},
    \mathbf{Q}_{k},
    \boldsymbol{\rho}(k),
    a_k,
    b_k,
    \mathbf{o}_{k}
    \right),
    \label{eq:memory_entry}
\end{equation}
where $\boldsymbol{\phi}_{k}$ is a compact DT fingerprint, $\boldsymbol{\rho}(k)$ is the RAT ranking produced by the active NS branch, $a_k$ is the selected RAT, $b_k\in\{\mathrm{madm},\mathrm{llm}\}$ identifies the branch that produced the decision, and $\mathbf{o}_{k}$ contains decision-quality indicators.

The fingerprint $\boldsymbol{\phi}_{k}$ is constructed by concatenating encoded service and preference indicators with normalized summary statistics of $\mathbf{H}_{k}$ and $\mathbf{Q}_{k}$. It is used only for similarity-based retrieval from DT memory and does not replace current-state evaluation. The memory is updated as
\begin{equation}
    \mathcal{M}_{k+1}
    =
    \operatorname{clip}_{M_{\max}}
    \left(
    \mathcal{M}_{k}\cup\{m_k\}
    \right),
    \label{eq:memory_update}
\end{equation}
where $M_{\max}$ is the maximum memory size. The DT evolves through UE-position updates, candidate-set changes, channel variation, packet-level dynamics, service/preference updates, and memory updates. This evolution is written as
\begin{equation}
    \mathcal{T}_{k+1}
    =
    \Psi
    \left(
    \mathcal{T}_{k},
    \boldsymbol{\Delta}_{k+1},
    \mathcal{M}_{k+1}
    \right),
    \label{eq:dt_evolution}
\end{equation}
where $\boldsymbol{\Delta}_{k+1}$ collects the updated UE position, candidate-set, channel, QoS, service, and preference components.

\subsection{DT-Grounded Network Selection Problem}
\label{subsec:dt_ns_problem}

The DT-grounded NS policy maps the decision context to a RAT ranking and a selected RAT, as follows
\begin{equation}
    \left(
    \boldsymbol{\rho}(k),
    a_k
    \right)
    =
    \pi
    \left(
    \mathcal{Z}_{k}
    \right),
    \qquad
    a_k\in\mathcal{A}(k).
    \label{eq:policy_mapping}
\end{equation}
The generic notation $\boldsymbol{\rho}(k)$ denotes the ranking produced by the selected NS branch. In this paper, $\pi(\cdot)$ is instantiated through two complementary branches. The first branch is MADM--LLM--NS. In this branch, the UIA provides the unified requirement profile $\mathbf{q}(k)$, from which the criterion weights and directions are extracted. A deterministic MADM operator then ranks the candidate RATs using the DT-derived decision matrix,
\begin{equation}
    \boldsymbol{\rho}_{\mathrm{madm}}(k)
    =
    f_{\mathrm{madm}}
    \left(
    \mathbf{DM}(k),
    \mathbf{q}(k),
    \mathcal{M}_{k}
    \right),
    \label{eq:madm_llm_ranking}
\end{equation}
where $f_{\mathrm{madm}}(\cdot)$ denotes the LLM-assisted MADM mapping. The memory $\mathcal{M}_{k}$ is used to support history-aware normalization, while the final ordering is produced by the selected MADM operator, such as SAW or TOPSIS. The selected RAT is
\begin{equation}
    a_{\mathrm{madm}}(k)
    =
    \rho_{\mathrm{madm},1}(k),
    \label{eq:madm_llm_selection}
\end{equation}
where $\rho_{\mathrm{madm},1}(k)$ is the first element of the MADM--LLM--NS ranking.

The second branch is the direct LLM--NS branch. In this branch, the LLM directly performs structured listwise ranking over the DT-derived candidate context,
\begin{equation}
    \boldsymbol{\rho}_{\mathrm{llm}}(k)
    =
    f_{\mathrm{llm}}
    \left(
    \mathbf{DM}(k),
    \mathbf{H}_{k},
    \mathbf{q}(k),
    \mathcal{M}_{k}^{\mathrm{ret}}
    \right),
    \label{eq:llm_ns_ranking}
\end{equation}
where $f_{\mathrm{llm}}(\cdot)$ denotes the direct LLM-based ranking function and $\mathcal{M}_{k}^{\mathrm{ret}}\subseteq\mathcal{M}_{k}$ denotes the RA--ICL retrieval set, i.e., similar historical DT-memory entries injected as in-context examples. Unlike the MADM--LLM--NS branch, this branch does not compute an explicit MADM score. The selected RAT is
\begin{equation}
    a_{\mathrm{llm}}(k)
    =
    \rho_{\mathrm{llm},1}(k),
    \label{eq:llm_ns_selection}
\end{equation}
where $\rho_{\mathrm{llm},1}(k)$ is the first element of the direct LLM--NS ranking. To quantify service-aware decision quality, let $x_{k,j}^{(r)}$ denote the value of criterion $c_j$ for RAT $r$. The criterion-level satisfaction score is
\begin{equation}
\psi_{k,j}^{(r)} =
\begin{cases}
\min\!\left\{x_{k,j}^{(r)}/(\theta_j^{\mathbf{S}(k)}+\epsilon),\,1\right\}, 
& c_j\!\in\!\mathcal{C}^{+},\\
\min\!\left\{\theta_j^{\mathbf{S}(k)}/(x_{k,j}^{(r)}+\epsilon),\,1\right\}, 
& c_j\!\in\!\mathcal{C}^{-}.
\end{cases}
\label{eq:criterion_satisfaction}
\end{equation}
The service-aware QoS satisfaction score is then
\begin{equation}
    \Gamma_{k}^{(r)}
    =
    \sum\nolimits_{j=1}^{|\mathcal{C}|}
    w_j(k)
    \psi_{k,j}^{(r)}.
    \label{eq:qos_satisfaction}
\end{equation}

A HO is unnecessary if the selected RAT changes while the previous RAT remains available and the new RAT does not provide a sufficient service-level improvement. For a margin $\Delta_{\mathrm{ho}}>0$,
\begin{equation}
\begin{aligned}
H_k^{\mathrm{UHO}}
=
\mathbb{I}\Big\{
& a_k\neq a_{k-1},\;
a_{k-1}\in\mathcal{A}(k), \\
& \Gamma_{k}^{(a_k)}
-
\Gamma_{k}^{(a_{k-1})}
<
\Delta_{\mathrm{ho}}
\Big\}.
\end{aligned}
\label{eq:uho_indicator}
\end{equation}

Ranking stability is evaluated under candidate-set evolution. Let $\mathcal{A}'(k)$ denote a perturbed candidate set and let $\mathcal{A}^{\cap}(k)=\mathcal{A}(k)\cap\mathcal{A}'(k)$ be the retained set. A rank-reversal event occurs if the relative order of any two retained RATs changes:
\begin{equation}
    I_k^{\mathrm{RR}}
    =
    \mathbb{I}
    \left\{
    \exists r_i,r_j\in\mathcal{A}^{\cap}(k):
    \Delta_{ij}(k)\Delta'_{ij}(k)<0
    \right\},
    \label{eq:rr_indicator}
\end{equation}
where $\Delta_{ij}(k)=\operatorname{pos}_{\boldsymbol{\rho}(k)}(r_i)-\operatorname{pos}_{\boldsymbol{\rho}(k)}(r_j)$, and $\Delta'_{ij}(k)$ is defined similarly after perturbation. The design goal is to improve service-aware QoS satisfaction while reducing unnecessary HOs and rank instability. Accordingly, the proposed framework is evaluated using $\Gamma_k^{(a_k)}$, $H_k^{\mathrm{VHO}}$, $H_k^{\mathrm{UHO}}$, and $I_k^{\mathrm{RR}}$.



\section{Physical-to-Digital Twin Construction}
\label{sec:physical_to_dt}

This section presents the first layer of the proposed framework on physical-to-DT construction. It builds the site-specific and geometry-aware representation from which the DT channel state ($\mathbf{H}_k$) is derived. This layer provides the physical and radio-environment basis for cross-layer synchronization and DT-grounded decision intelligence.

\subsection{Site-Specific 3D Scene and Scenario Definition}
\label{subsec:geospatial_twin}

The physical-to-DT construction starts from a geospatial description of the service area. OpenStreetMap (OSM) is used to extract the main map elements, including building footprints, roads, and street layout~\cite{haklay2008openstreetmap}. The extracted data are then imported into Blender to generate the corresponding site-specific 3D scene, defining the environment component \(\mathcal{E}\).


To enable geometry-aware propagation modeling, the main objects are assigned radio-material labels, such as concrete, glass, metal, asphalt, foliage, and ground surfaces. The resulting material-aware 3D scene is then used by the ray-tracing engine for channel-descriptor extraction. The wireless scenario is constructed by placing RAT-specific access nodes, assigning their radio configurations, and mapping the UE trajectory onto the same coordinate system. Each access node follows the RAT configuration \(\boldsymbol{\eta}_{r}\) in~\eqref{eq:rat_config}, including carrier frequency, bandwidth, noise figure, transmit power, antenna configuration, and node position. At each decision epoch, the scenario provides the UE position \(\mathbf{p}_k\), while coverage and link-feasibility conditions define the candidate set \(\mathcal{A}(k)\). The service profile \(\mathbf{S}(k)\) and user/operator preference context \(\mathbf{U}(k)\) are retained in the DT state and later translated by the UIA into the requirement profile \(\mathbf{q}(k)\).


\subsection{Geometry-Aware Ray-Tracing Twin Engine}
\label{subsec:ray_tracing_engine}

The geometry-aware ray-tracing engine maps the material-aware 3D scene, RAT deployment, and UE position into compact radio descriptors. In the implementation, this mapping is realized using Sionna RT with the Mitsuba backend. For each epoch \(k\) and RAT \(r\), the ray-tracing stage evaluates the site-specific propagation between the transmitter, the UE position, and the surrounding environment.

The output is the channel descriptor \(\mathbf{h}_{k}^{(r)}\) in~\eqref{eq:channel_descriptor}, which summarizes the propagation state through channel gain, first-arrival delay, RMS delay spread, and the number of valid propagation paths. Sionna RT is used only to construct the geometry-aware radio component of the DT, while received power, SINR, and packet-level QoS are derived during cross-layer synchronization. The descriptors of all feasible RATs are collected into \(\mathbf{H}_k\), which captures the radio-environment state of the candidate RATs and is retained in the DT-derived decision context \(\mathcal{Z}_k\).


\section{Cross-Layer Synchronization}
\label{sec:cross_layer_sync}

After constructing the DT, cross-layer synchronization maps the geometry-aware radio descriptors to packet-level emulation to obtain the service-level QoS evidence used by the NS policy.




\begingroup
\color{black}

\subsection{ns-3 Packet-Level DT Emulation}
\label{subsec:ns3_packet_emulation}

For each decision epoch $k$ and candidate RAT $r$, the physical-to-DT layer provides a geometry-aware channel descriptor obtained from Sionna~RT. This descriptor contains the ray-traced channel gain $G_k^{(r)}$, first-arrival delay $\tau_k^{(r)}$, RMS delay spread $\sigma_{\tau,k}^{(r)}$, and the number of valid propagation paths $L_k^{(r)}$, as defined in~\eqref{eq:channel_descriptor}. These quantities are not treated as packet-level QoS metrics; rather, they provide the radio-environment evidence from which the packet-level emulator derives link behavior. The RAT configuration, including carrier frequency, bandwidth, transmit power, antenna model, and noise figure, is jointly used with the channel descriptor during the Sionna~RT--to--ns-3 synchronization stage.

The synchronization is performed at each decision epoch by mapping the ray-traced channel descriptor to link-level quantities used by ns-3. For each RAT, the serving access site is selected according to the strongest ray-traced channel gain. The received power and SINR are then derived from the selected gain, the RAT transmit power, antenna gain, bandwidth, and noise figure. In linear scale, the SINR can be written as
\begin{equation}
\mathrm{SINR}_k^{(r)} = \frac{P_{k,\mathrm{rx}}^{(r)}}{N_0 W_r F_r + I_k^{(r)}},
\label{eq:sinr_sync}
\end{equation}

where $P_{k,\mathrm{rx}}^{(r)}$ is the received power of RAT $r$ at epoch $k$, $N_0$ denotes the thermal-noise power spectral density in linear units, $W_r$ is the RAT bandwidth in Hz, $F_r$ is the noise figure converted to linear scale, and $I_k^{(r)}$ denotes co-channel interference. In this evaluation, a noise-limited operating point is considered, i.e., $I_k^{(r)}\to 0$. Therefore, the noise term is computed as $N_0 W_r F_r$, with all dB-valued quantities converted to linear scale before the SINR calculation.

The resulting SINR is used by the packet-level link abstraction to determine the effective spectral efficiency, physical-layer rate, and packet-error behavior configured in ns-3. In the implementation, the SINR is mapped to a discrete modulation-and-coding scheme (MCS) level using a gap-adjusted Shannon expression with RAT-specific overhead and maximum spectral-efficiency limits. The packet-error process is then configured through a smooth block-error-rate (BLER) curve whose transition width depends on the RMS delay spread. The highest supported level satisfying the target BLER constraint is retained. The objective of this abstraction is not to emulate bit-level physical-layer processing, but to provide a consistent packet-level interface between the geometry-aware radio state and the network-level QoS indicators. The selected rate, propagation delay, and packet-error process are then used to configure a packet-level ns-3 run for each candidate RAT.

The ns-3 emulator accounts for packet generation, serialization, queueing, propagation delay, packet reception, and packet loss over a fixed observation window $T_{\mathrm{obs}}$. The QoS indicators are measured from the transmitted and received packet traces. Let $\mathcal{P}_{k,r}^{\mathrm{tx}}$ and $\mathcal{P}_{k,r}^{\mathrm{rx}}$ denote the sets of packets transmitted and successfully received within the observation window for RAT $r$ at epoch $k$. For each received packet $p$, let $t_p^{\mathrm{tx}}$ and $t_p^{\mathrm{rx}}$ denote its generation and reception times, and let $d_p = t_p^{\mathrm{rx}} - t_p^{\mathrm{tx}}$ be the packet delay. The mean delay is computed as
\begin{equation}
D_k^{(r)} = \frac{1}{|\mathcal{P}_{k,r}^{\mathrm{rx}}|} \sum_{p \in \mathcal{P}_{k,r}^{\mathrm{rx}}} d_p .
\label{eq:delay_trace}
\end{equation}
Ordering the received packets by reception time as $p_1,\ldots,p_M$, with $M = |\mathcal{P}_{k,r}^{\mathrm{rx}}|$, the jitter is computed as the mean absolute variation of consecutive packet delays,
\begin{equation}
J_k^{(r)} = \frac{1}{M-1} \sum_{i=2}^{M} \left| d_{p_i} - d_{p_{i-1}} \right| .
\label{eq:jitter_trace}
\end{equation}
When fewer than two packets are received, the jitter is set to zero. The packet-loss ratio is computed from the transmitted and received packet counts as
\begin{equation}
\mathrm{PLR}_k^{(r)} = 1 - \frac{|\mathcal{P}_{k,r}^{\mathrm{rx}}|}{|\mathcal{P}_{k,r}^{\mathrm{tx}}|},
\label{eq:plr_trace}
\end{equation}
and therefore captures both queue-overflow losses and link-error losses. The throughput is computed from the correctly received bytes over the observation window. EE is computed in post-processing from the correctly delivered bits and the adopted per-RAT power-consumption model, following the bits-per-Joule convention~\cite{buzzi2016survey,mefgouda2024qos}, while the access cost $C_k^{(r)}$ is the static per-RAT score reported in Table~\ref{tab:rat_config}. These quantities form the QoS vector $\mathbf{x}_{k}^{(r)}$ in~\eqref{eq:qos_vector}.


A common packet-generation process is used across all compared NS schemes to ensure a controlled comparison. The service profile does not modify the packet generator in the reported evaluation; instead, it determines the decision priorities and QoS acceptability targets used by the UIA, MADM--LLM--NS, LLM--NS, and the satisfaction evaluation. This separation ensures that the observed performance differences are caused by the NS decision logic rather than by different traffic realizations. 

\endgroup

\subsection{Multi-RAT QoS Synchronization and Decision Context}
\label{subsec:multi_rat_qos_sync}

All candidate RATs are evaluated under the same UE state, service profile, and observation interval. Thus, the candidates in $\mathcal{A}(k)$ are compared under a common DT context rather than under independent emulation conditions. At epoch $k$, each feasible RAT produces one QoS vector, and these vectors form the DT-derived decision matrix $\mathbf{DM}(k)$ in~\eqref{eq:decision_matrix}.

The candidate set \(\mathcal{A}(k)\) varies with the UE position, blockage, coverage conditions, and link-feasibility constraints. This evolution is retained because candidate insertion and removal directly affect RRP robustness and VHO stability. After channel and QoS generation, the framework synchronizes the radio-environment, packet-level, service-intent, and memory components into a unified decision context for the LLM-based intelligence layer. The ray-tracing engine does not evaluate service satisfaction, and the packet-level emulator does not select a RAT; both provide evidence for the subsequent DT-grounded NS decision.

\section{LLM-Based Digital Twin Intelligence}
\label{sec:llm_dt_intelligence}

The LLM is introduced as a DT-grounded intelligence layer operating over the DT-derived decision context, the active service/user intent, and the DT memory. Rather than acting as a generic ranker, it produces a structured requirement profile, a RAT ranking, and the corresponding access decision. This design follows the emerging role of LLMs in telecom systems as intent-interpretation and decision-support modules rather than standalone wireless models~\cite{Zhou2024LLMTelecomSurvey}.


\subsection{Unified Intent Agent}
\label{subsec:uia}

The UIA implements the intent-to-requirement mapping defined in~\eqref{eq:req_profile}--\eqref{eq:uia_weight_sum}. It receives the user/operator preference context $\mathbf{U}(k)$, the active service profile $\mathbf{S}(k)$, the criterion set $\mathcal{C}$, and, when enabled, a compact summary of the current candidate landscape. Its output is the unified requirement profile $\mathbf{q}(k)$, which specifies the criterion type, normalized weight, qualitative importance, and criterion-level justification. The UIA does not rank candidate RATs and does not output the final access decision. It only converts high-level service/user intent into machine-readable decision semantics consumed by the downstream NS branches.


As shown in Fig.~\ref{fig:uia_prompt}, the UIA prompt is constrained to return a valid structured profile over the predefined NS criteria. It is not allowed to introduce new criteria, candidate-specific utilities, thresholds, RAT rankings, or selected RATs. This constraint prevents the intent agent from becoming an implicit ranker: the LLM first resolves the active service/user intent into structured decision priorities, and the resulting profile $\mathbf{q}(k)$ is then shared by both branches. The MADM--LLM--NS branch uses its criterion directions and weights, whereas the direct LLM--NS branch uses the full profile together with the DT-derived candidate context.


\begin{figure}[!htbp]
\centering
\small
\setlength{\fboxsep}{4pt}
\fbox{%
\begin{minipage}{\dimexpr\linewidth-2\fboxsep-2\fboxrule\relax}

\raggedright
\textbf{Input:} user/operator preference $\mathbf{U}(k)$, service profile $\mathbf{S}(k)$, criterion set $\mathcal{C}$, and optional candidate summary $\boldsymbol{\zeta}(k)$.

\textbf{Task:} infer how the active user/service intent should prioritize the NS criteria.

\textbf{Output:} for each criterion, return \texttt{\{type, weight, importance, justification\}}.

\textbf{Rules:} use only the predefined NS criteria; assign non-negative weights; do not output thresholds, candidate utilities, RAT rankings, or a selected RAT; return valid JSON only.
\end{minipage}%
}
\caption{Compact UIA prompt schema.}
\label{fig:uia_prompt}
\end{figure}


\subsection{MADM--LLM--NS Branch}
\label{subsec:madm_llm_ns}

The MADM--LLM--NS branch preserves the deterministic structure of classical MADM-based NS while replacing the weighting technique with the UIA-derived requirement profile.  The LLM does not select the RAT in this branch; it only provides service-aware criterion priorities. The final ordering is governed by a standard MADM operator, either TOPSIS or SAW, representing distance-to-ideal ranking and direct weighted aggregation, respectively~\cite{khalili2025tvt,dos2025network,ndashimye2021multi}.

To reduce normalization-induced candidate-set sensitivity, the branch uses HAAN. The motivation is the RRP, in which the relative order of retained alternatives changes after candidate insertion or removal even though the retained candidates are themselves unchanged~\cite{jahan2015state,tu2025analytic}. In MADM--NS, this instability originates mainly at the normalization stage, because candidate insertion or removal alters the per-epoch normalization references and therefore the normalized scores of retained RATs~\cite{senouci2016topsis,baghla2018vikor,mefgouda2024qos}. HAAN addresses this by estimating normalization references from recent DT memory rather than from the instantaneous candidate set alone.

For criterion $c_j$, the HAAN reference set is
\begin{equation}
    \mathcal{V}_{k,j}^{\mathrm{HAAN}}
    =
    \{x_{k,j}^{(r)}:r\in\mathcal{A}(k)\}
    \cup
    \{x_{i,j}^{(r)}:m_i\in\mathcal{M}_{k}^{\mathrm{rel}},\,r\in\mathcal{A}(i)\},
    \label{eq:haan_reference_set}
\end{equation}
where $\mathcal{M}_{k}^{\mathrm{rel}}\subseteq\mathcal{M}_k$ denotes recent memory entries consistent with the active service profile. The resulting references keep the MADM branch deterministic while making the normalized values of retained candidates less sensitive to candidate-set evolution.

Let $\bar{x}_{k,j}^{(r)}$ denote the direction-aligned HAAN-normalized value of criterion $c_j$ for RAT $r$, where larger values are preferred after benefit/cost handling. For SAW--LLM--NS, the score is
\begin{equation}
    s_{\mathrm{SAW}}^{(r)}(k)
    =
    \sum\nolimits_{j=1}^{|\mathcal{C}|}
    w_j(k)\,\bar{x}_{k,j}^{(r)}.
    \label{eq:saw_llm_score}
\end{equation}
For TOPSIS--LLM--NS, the weighted normalized value is $v_{k,j}^{(r)}=w_j(k)\bar{x}_{k,j}^{(r)}$, and the closeness score is
\begin{equation}
    s_{\mathrm{TOPSIS}}^{(r)}(k)
    =
    \frac{d_{r}^{-}(k)}{d_{r}^{+}(k)+d_{r}^{-}(k)},
    \label{eq:topsis_llm_score}
\end{equation}
where $d_{r}^{+}(k)$ and $d_{r}^{-}(k)$ are the Euclidean distances of RAT $r$ to the ideal and anti-ideal alternatives in the weighted normalized space. Candidates are ordered in descending score.

\subsection{Direct LLM--NS Branch}
\label{subsec:direct_llm_ns}

The direct LLM--NS branch performs structured listwise ranking over the DT-derived candidate context. As defined in~\eqref{eq:llm_ns_ranking}, its input is $(\mathbf{DM}(k),\mathbf{H}_k,\mathbf{q}(k),\mathcal{M}_{k}^{\mathrm{ret}})$. Unlike the MADM--LLM--NS branch, it does not compute a scalar MADM score; it directly returns an ordered RAT list and the selected RAT.

The prompt schema is summarized in Fig.~\ref{fig:llm_ns_prompt}. The model is instructed to respect benefit/cost directions, follow the priority order encoded in $\mathbf{q}(k)$, base the ranking only on measurable DT evidence, and emit a justification used for auditability rather than as a substitute for networking-level evaluation.

\begin{figure}[!htbp]
\centering
\small
\setlength{\fboxsep}{4pt}
\fbox{%
\begin{minipage}{\dimexpr\linewidth-2\fboxsep-2\fboxrule\relax}
\raggedright
\textbf{Input:} candidate RAT table $\mathbf{DM}(k)$, compact channel descriptors from $\mathbf{H}_k$, requirement profile $\mathbf{q}(k)$, service-specific few-shot examples, and retrieved DT-memory entries $\mathcal{M}_{k}^{\mathrm{ret}}$.

\textbf{Task:} rank all candidate RATs from most suitable to least suitable for the active service/user context.

\textbf{Rules:} respect benefit/cost directions; prioritize criteria according to $\mathbf{q}(k)$; use retrieved memory as evaluated precedent, not as a hard label; base the ranking only on measurable DT evidence.

\textbf{Output:} valid JSON containing the ordered RAT list, selected RAT, acceptability flag, and concise service-aware justification.
\end{minipage}
}
\caption{Compact prompt schema used by the direct LLM--NS branch.}
\label{fig:llm_ns_prompt}

\end{figure}

The justification is used for auditability and consistency checking; it is not used as a substitute for the networking-level evaluation.

\noindent\textbf{Service-specific few-shot demonstration.}
When enabled, the prompt is augmented with compact in-context demonstrations that show how the requirement profile should be applied to candidate landscapes. A representative VR/AR example is given in Fig.~\ref{fig:fewshot_vrar_example}, where the model is guided to jointly prioritize PLR, delay, and jitter rather than to select a RAT based on a single dominant metric.

\begin{figure}[!htbp]
\centering
\setlength{\fboxsep}{4pt}
\fbox{%
\begin{minipage}{\dimexpr\linewidth-2\fboxsep-2\fboxrule\relax}
\small
\raggedright
\textbf{Example: VR/AR few-shot demonstration.}

\medskip
\textbf{Requirement profile:} PLR is critical; delay and jitter are high priority; EE and cost are secondary.

\medskip
\textbf{Candidate context:}\\

Candidate $A$: PLR $=0.08\%$, $D=0.09$ ms, $J=0.015$ ms, EE $=7.1{\times}10^{5}$ bit/J, $C= 7$;\\
Candidate $B$: PLR $=0.01\%$, $D=0.22$ ms, $J=0.036$ ms, EE $=7.3{\times}10^{6}$ bit/J, $C=1$;\\
Candidate $C$: PLR $=1.30\%$, $D=0.11$ ms, $J=0.035$ ms, EE $=1.6{\times}10^{7}$ bit/J, $C=2$.
\medskip
\raggedright
\textbf{Reference ranking:} $A > B > C$.
\\
\textbf{Decision pattern:} $A$ is preferred because it jointly satisfies the dominant VR/AR criteria. $B$ has lower PLR and lower cost but worse delay and jitter. $C$ is rejected because its PLR violates the reliability requirement.
\end{minipage}
}
\caption{Representative few-shot in-context demonstration used by the direct LLM--NS branch for the VR/AR service profile.}
\label{fig:fewshot_vrar_example}
\end{figure}

\subsection{DT Memory and IIA-Constrained RA--ICL}
\label{subsec:rag_icl}

The direct LLM--NS branch is stabilized through a RA--ICL mechanism that operates over the DT memory $\mathcal{M}_k$ and incorporates IIA-inspired retained-order preservation under candidate-set evolution~\cite{lewis2020retrieval,brown2020language}. At each decision epoch, the current candidate set $\mathcal{A}(k)$ is compared, under the active service profile $\mathbf{S}(k)$, against the candidate sets of past memory entries, and one of four cases is executed.

\paragraph{Cold start} If no past entry under $\mathbf{S}(k)$ has a candidate set sufficiently similar to $\mathcal{A}(k)$, or if the QoS values of the retained candidates have drifted beyond a tolerance threshold, the LLM is invoked with the full prompt of Section~\ref{subsec:direct_llm_ns} and returns a complete listwise ranking $\boldsymbol{\rho}_{\mathrm{llm}}(k)$ over $\mathcal{A}(k)$. The context $(\mathbf{DM}(k),\boldsymbol{\rho}_{\mathrm{llm}}(k))$ is recorded in $\mathcal{M}_{k+1}$ for subsequent reuse.

\paragraph{Candidate withdrawal} If a retrieved past entry $m_i$ satisfies $\mathcal{A}(k)\subset\mathcal{A}(i)$, i.e., one or more RATs have left the candidate set since epoch $i$, the new ranking is obtained by projecting the past ranking onto the retained candidates:
\begin{equation}
    \boldsymbol{\rho}_{\mathrm{llm}}(k)
    =
    \operatorname{Proj}_{\mathcal{A}(k)}
    \bigl(\boldsymbol{\rho}_{\mathrm{llm}}(i)\bigr),
    \label{eq:rag_remove}
\end{equation}
where $\operatorname{Proj}_{\mathcal{A}(k)}(\cdot)$ removes the withdrawn RATs while preserving the relative order of the retained ones. The LLM is not invoked in this case.

\paragraph{Candidate insertion}
\begingroup
\color{black}
If $\mathcal{A}(i)\subset\mathcal{A}(k)$, i.e., one or more RATs have joined the candidate set, the retrieved ranking $\boldsymbol{\rho}_{\mathrm{llm}}(i)$ is used as an evaluated precedent rather than as an immutable ordering constraint. The LLM is invoked over the enlarged candidate context and receives the retained ranking, the added RATs, the current DT-derived decision matrix, the channel descriptors, and the requirement profile. The insertion-aware ranking is written as
\begin{equation}
\boldsymbol{\rho}_{\mathrm{llm}}(k)
=
f_{\mathrm{ins}}
\left(
\boldsymbol{\rho}_{\mathrm{llm}}(i),
\mathbf{DM}(k),
\mathbf{H}_{k},
\mathbf{q}(k),
\mathcal{A}(k)\setminus\mathcal{A}(i)
\right),
\label{eq:rag_add}
\end{equation}
where $f_{\mathrm{ins}}(\cdot)$ denotes the insertion-aware LLM ranking function. The retrieved order guides the decision and discourages unnecessary reordering, but the LLM is allowed to revise the enlarged ranking when the newly observed RAT introduces relevant DT evidence. Therefore, candidate insertion is not treated as a zero-RRP structural guarantee.

\paragraph{Mixed case}
If the candidate set both gains and loses RATs, equation \eqref{eq:rag_remove} is first applied to remove withdrawn RATs and preserve the ordering of the retained candidates. The insertion-aware ranking in~\eqref{eq:rag_add} is then applied to evaluate newly observed RATs in the enlarged candidate context.

This construction provides structural retained-order preservation under candidate withdrawal. In this case, the projection in~\eqref{eq:rag_remove} removes unavailable RATs while preserving the relative order of the retained candidates; hence, the retained-candidate rank-reversal indicator in~\eqref{eq:rr_indicator} is zero by construction. Candidate insertion is treated differently because new RAT evidence is introduced into the decision context. The retrieved ranking is used as an evaluated precedent, but the LLM is allowed to evaluate the newly observed RAT within the enlarged candidate context through~\eqref{eq:rag_add}. As a result, insertion may produce residual rank reversals when the new evidence changes the relative preference structure. This behavior reflects the intended stability--adaptivity balance of RA--ICL: withdrawal is handled by structural preservation, whereas insertion allows adaptation to newly observed RAT evidence. The cold-start rule remains a safety mechanism when the candidate set or the QoS evidence of retained candidates drifts beyond tolerance.

\noindent\textbf{Stability--optimality trade-off.}
The proposed RA--ICL mechanism should be interpreted as preserving retained-candidate order under withdrawal and as constraining, but not eliminating, ranking changes under insertion. It does not assert that the resulting ranking is globally optimal for every evolved candidate set. Under withdrawal, no new candidate evidence is introduced, and projection preserves the retained ordering for stability. Under insertion, new evidence is introduced, and stability may conflict with adaptivity: preserving the previous retained order favors stable mobility behavior, whereas re-evaluating the enlarged candidate context may justify changing the ordering. This trade-off is deliberate, because frequent reordering in mobility-driven NS can trigger unnecessary VHOs and ping-pong effects. The cold-start path acts as a safety mechanism: whenever the candidate set or the QoS evidence of retained candidates drifts beyond tolerance, the full listwise LLM ranking is recomputed, allowing new evidence to update the ranking.




\endgroup

\section{Evaluation and Memory Update}
\label{sec:evaluation_memory_update}

Finally, the proposed system evaluates the selected RAT, records the decision outcome, and updates the DT memory for subsequent epochs. After an NS branch produces the ranking and selected RAT, the evaluation layer computes the service-aware satisfaction score, VHO indicator, unnecessary-HO indicator, and RRP indicator defined in Section~\ref{sec:system_model}. These metrics assess service satisfaction, HO necessity, and ranking stability under candidate-set evolution.


After evaluation, the DT memory is updated using the memory-entry structure defined in~\eqref{eq:memory_entry}. The updated memory is not used as a cache of past decisions; instead, it stores evaluated DT contexts and outcomes that can later support HAAN-based normalization in the MADM--LLM--NS branch and RA--ICL retrieval in the direct LLM--NS branch.  Thus, this layer makes the DT an evolving decision-support representation rather than a one-shot simulation dataset. Each selected RAT is evaluated, stored with its DT context, and reused to improve normalization stability and memory-assisted reasoning in subsequent decision epochs.



\section{Simulation Setup \& Evaluation}
\label{sec:eval_setup}
This section describes the simulation setup and evaluation protocol to assess the proposed DT-grounded NS framework.

\subsection{Wireless DT Realization}
\label{subsec:sim_scenario}
\begin{figure*}[!htbp]
    \centering
    \subfloat[\scriptsize Site-specific 3D DT representation.]{
        \includegraphics[width=0.42\linewidth]{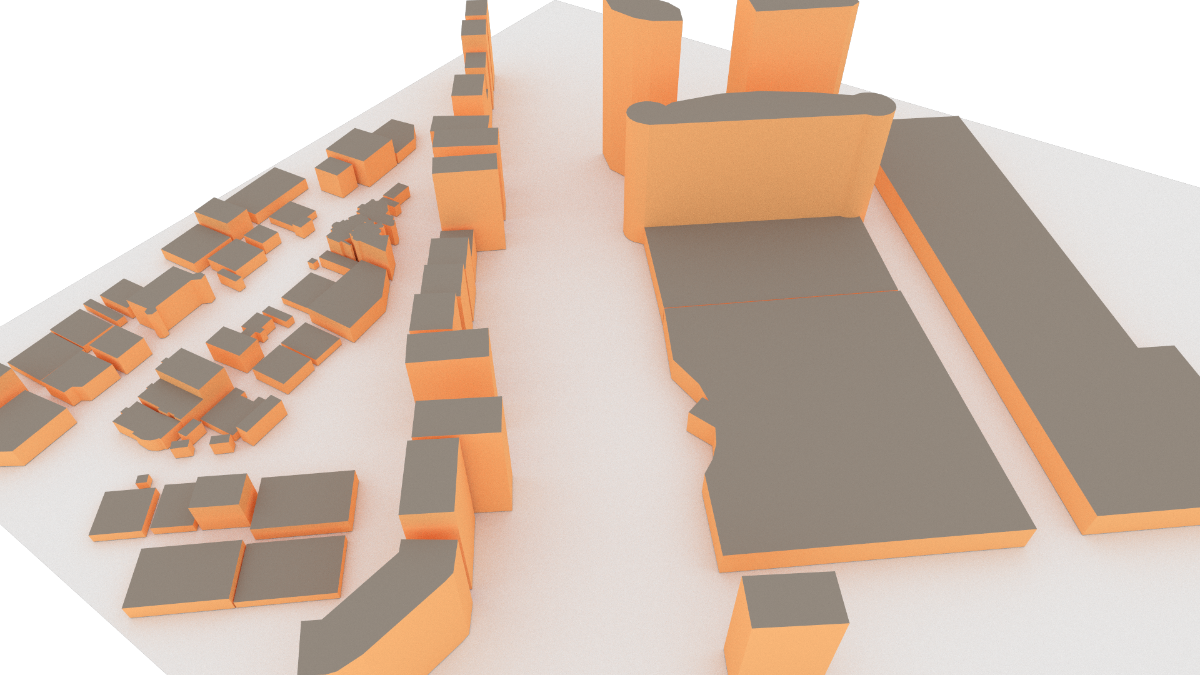}
        \label{fig:dt_scene_only}
    }
     \hspace{2mm}
    \subfloat[\scriptsize Representative geometry-aware propagation state.]{
        \includegraphics[width=0.42\linewidth]{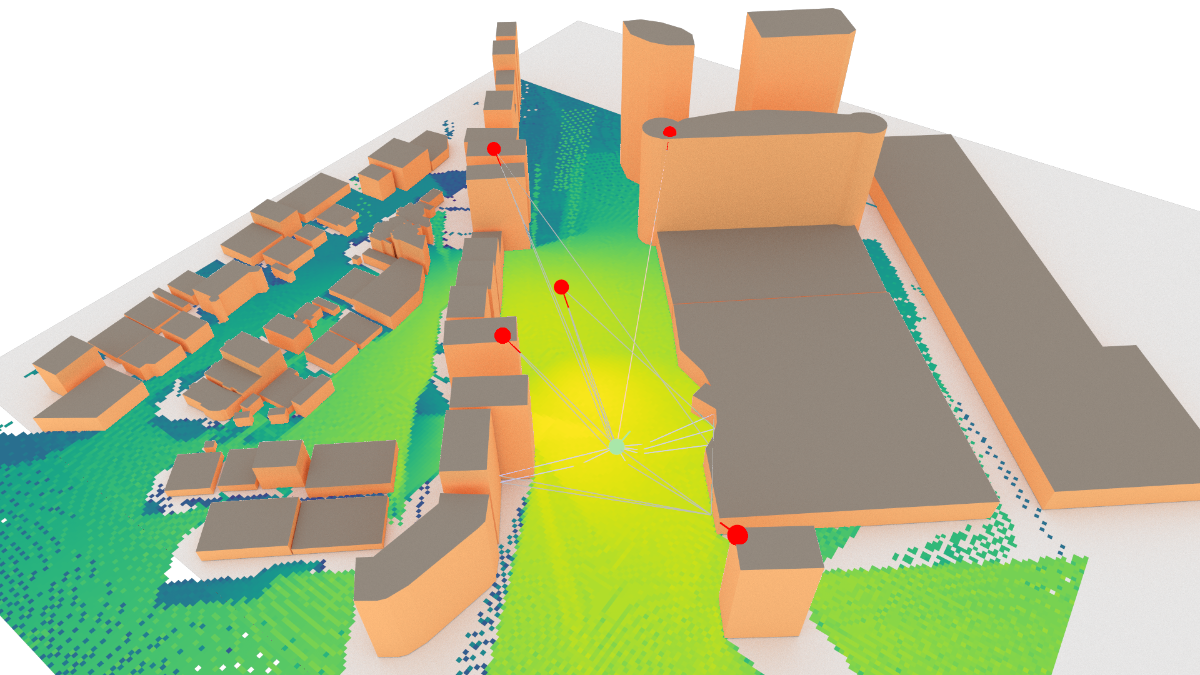}
        \label{fig:dt_rt_paths}
    }
    \caption{Site-specific wireless DT representation and geometry-aware propagation state.}
    \label{fig:dt_scenario}
    \vspace{-2mm}
\end{figure*}

The evaluation is conducted over a site-specific urban region in Abu Dhabi, UAE. The 3D environment is extracted from OSM and refined in Blender, then passed to Sionna RT to obtain geometry-aware channel descriptors, which are subsequently converted into packet-level QoS indicators by ns-3. Fig.~\ref{fig:dt_scenario} shows the considered scene and a representative ray-tracing output.

A trajectory of $K=2000$ UE positions is used to evolve the DT state. All compared schemes operate over the same DT trace, candidate RATs, and service profiles, so any performance gap reflects the decision logic rather than the underlying radio or traffic realization. The DT supports state representation, candidate-RAT evaluation, application-aware access selection, and memory-based decision reuse; physical-layer closed-loop control is outside the scope of this evaluation.



\subsection{Multi-RAT Deployment and Service Profiles}
\label{subsec:rat_deployment}

The HWN comprises five RATs, listed in Table~\ref{tab:rat_config}: WiMAX, Wi-Fi, Wi-Fi~6, LTE, and 5G~NR. Their parameters instantiate the configuration vector in~\eqref{eq:rat_config}: carrier frequency, bandwidth, transmit power, antenna pattern, and noise figure are consumed by the Sionna~RT--to--ns-3 synchronization stage, while the cost score $C$ enters the DT-derived decision matrix as the access-cost criterion. Because the RATs differ across all of these parameters, no single radio metric determines the preferred RAT.

\begin{table}[!htbp]
\small
\caption{Per-RAT simulation configuration.}
\label{tab:rat_config}
\centering
\footnotesize
\setlength{\tabcolsep}{2pt}
\begin{tabular}{p{2.1cm}|c|c|c|c|c}
\hline
\textbf{Parameter} & \textbf{WiMAX} & \textbf{Wi-Fi} & \textbf{Wi-Fi~6} & \textbf{LTE} & \textbf{5G NR} \\
\hline
Carrier frequency (GHz) & 3.5 & 2.4 & 5.5 & 1.8 & 3.5 \\
Bandwidth (MHz)         & 10  & 20  & 80  & 20  & 100 \\
Tx power (dBm)          & 30  & 20  & 23  & 46  & 43  \\
Noise figure (dB)       & 7   & 6   & 6   & 7   & 7  \\
Antenna pattern         & TR~38.901 & Isotropic & Isotropic & TR~38.901 & TR~38.901 \\
Cost score $C$          & 4 & 1 & 2 & 3 & 7 \\
\hline
\end{tabular}
\end{table}
In this work, each candidate RAT is evaluated over an observation window of $T_{\mathrm{obs}}=10$~s, and the packet-level emulation settings are summarized in Table~\ref{tab:emulation_settings}. The traffic source is a bursty User Datagram Protocol (UDP) generator controlled by a token-bucket process. The token-bucket burst cap is an implementation upper bound and does not correspond to a constant offered rate; instead, the actual transmitted traffic is constrained by the token-bucket state, effective link rate, queue occupancy, and packet-error process.
\begin{table}[!htbp]
\small
\caption{Packet-level emulation settings common to all compared schemes.}
\label{tab:emulation_settings}
\centering
\footnotesize
\setlength{\tabcolsep}{6pt}
\begin{tabular}{l|c}
\hline
\textbf{Parameter} & \textbf{Value} \\
\hline
Network emulator            & ns-3 \\
Traffic source              & Bursty UDP with token-bucket control \\
Packet size                 & 1200 B \\
Observation window $T_{\mathrm{obs}}$ & 10 s \\
Sender update interval      & 0.3 ms \\
Token-bucket burst cap      & 1200 packets/update \\
Link-abstraction target BLER & $\le 10\%$ \\
Queue discipline            & DropTail, 70--100 packets \\
\hline
\end{tabular}
\end{table}


Five service profiles are considered, as summarized in Table~\ref{tab:service_profiles}. Following the 3GPP QoS traffic-class taxonomy in TS~23.107~\cite{3gpp_ts_23_107_v18}, the profiles are grouped into conversational, streaming, interactive, and background classes. The considered NS criteria are delay, jitter, packet-loss rate, EE, and access cost, which are consistent with service-aware heterogeneous access selection in~\cite{zhu2021adaptive}. For each profile, the dominant criteria are selected according to the main QoS sensitivity of the corresponding traffic class. VR/AR is treated as an interactive service because immersive traffic is sensitive to reliability, latency, and delay variation; therefore, PLR, delay, and jitter are dominant. Industrial control is mapped to the conversational class, where fast response is critical, so delay is dominant, while access cost is also considered to reflect the resource cost of maintaining strict control connectivity. Office automation is treated as an interactive but less delay-critical service, where energy-efficient and reliable access is preferred; therefore, EE and PLR dominate. Video streaming belongs to the streaming class, where buffered delivery makes instantaneous delay and jitter less critical than delivery reliability and access cost; hence, cost and PLR are dominant. Background traffic is delay tolerant and mainly favors reliable low-cost delivery, so PLR and cost dominate. The same service-profile mapping and dominant-criterion assignment are used for all compared schemes to ensure a controlled evaluation.


\begin{table}[!htbp]
\caption{Service profiles and dominant decision criteria.}
\label{tab:service_profiles}
\vspace{-2mm}
\centering
\small
\setlength{\tabcolsep}{2pt}
\begin{tabular}{l|c|l}
\hline
\textbf{Service profile} & \textbf{3GPP-aligned class} & \textbf{Dominant criteria} \\
\hline
VR/AR & Interactive & PLR, delay, jitter \\
Industrial control & Conversational & Delay, cost \\
Office automation & Interactive & EE, PLR \\
Streaming & Streaming & Cost, PLR \\
Background & Background & PLR, cost \\
\hline
\end{tabular}
\end{table}

\subsection{Compared Schemes}
\label{subsec:baselines}
The compared schemes are as follows. TOPSIS--AHP and SAW--AHP are classical MADM--NS baselines using fixed AHP weights~\cite{vaidya2006analytic}; they implement distance-to-ideal ranking and compensatory weighted aggregation over the instantaneous decision matrix only, without access to the DT intelligence layer or the DT memory layer. TOPSIS--LLM--NS and SAW--LLM--NS exercise the MADM--LLM--NS branch of Section~\ref{subsec:madm_llm_ns}, in which the UIA of the DT intelligence layer supplies service-aware weights and HAAN draws normalization references from the DT memory layer. LLM--NS exercises the direct listwise branch of Section~\ref{subsec:direct_llm_ns} under the IIA-constrained RA--ICL mechanism of Section~\ref{subsec:rag_icl}, which operates over the same DT memory layer.  The AHP weights, computed using the standard pairwise-comparison procedure of AHP~\cite{vaidya2006analytic} as adopted for service-preference weighting in heterogeneous access selection~\cite{zhu2021adaptive}, are reported in Table~\ref{tab:ahp_weights}.

\begin{table}[!htbp]
\small
\caption{AHP-derived weights for baseline schemes.}
\label{tab:ahp_weights}
\centering
\small
\setlength{\tabcolsep}{7pt}
\begin{tabular}{l|c|c|c|c|c}
\hline
\textbf{Profile} & $w_{\mathrm{EE}}$ & $w_D$ & $w_{\mathrm{PLR}}$ & $w_J$ & $w_C$ \\
\hline
VR/AR      & 0.13 & 0.26 & 0.38 & 0.20 & 0.03 \\
Industrial & 0.10 & 0.45 & 0.20 & 0.15 & 0.10 \\
Office     & 0.40 & 0.15 & 0.22 & 0.10 & 0.13 \\
Streaming  & 0.02 & 0.10 & 0.17 & 0.06 & 0.65 \\
Background & 0.05 & 0.10 & 0.30 & 0.10 & 0.45 \\
\hline
\end{tabular}
\end{table}
\subsection{LLM and Memory Configuration}
\label{subsec:llm_memory}

The LLM-based schemes use Qwen2.5--14B--Instruct as a frozen backbone, served through vLLM on an NVIDIA A100 GPU under greedy decoding. No fine-tuning is performed, so every observed gain is attributable to the UIA, the prompt design, and the DT-memory mechanisms (HAAN and IIA-constrained RA--ICL). The prompt templates, service profiles, retrieval implementation, and evaluation scripts are publicly available\footnote{Source code: \url{https://github.com/BrahiM-Mefgouda/TwinRAT-Select}}. The main parameters are listed in Table~\ref{tab:llm_memory_config}.

\begin{table}[!htbp]
\small
\caption{LLM and memory configuration.}
\label{tab:llm_memory_config}
\centering
\vspace{-2mm}
\setlength{\tabcolsep}{1pt}
\begin{tabular}{l|c}
\hline
\textbf{Parameter} & \textbf{Value} \\
\hline
LLM backbone & Qwen2.5--14B--Instruct \\
Inference engine & vLLM \\
HAAN history window & 20 epochs \\
Stability constant $\epsilon$ & $10^{-6}$ \\
Few-shot demonstrations & 10 per service class \\
DT-memory size $M_{\max}$ & 200 \\
Retrieved entries $K_r$ & 3 \\
Decision epochs & 2000 \\
\hline
\end{tabular}
\end{table}

\subsection{Evaluation Protocol and Metrics}
\label{subsec:eval_metrics}


Four metrics are evaluated: RRP rate, QoS satisfaction, unnecessary VHO rate, and computational complexity. The RRP rate is reported across all five service profiles because it measures how sensitive each ranking method is to candidate-set perturbations. This allows the robustness of the selection logic to be evaluated independently of a single service profile. After this general RRP analysis, the detailed QoS satisfaction and unnecessary-VHO evaluations focus on the VR/AR profile. VR/AR is selected because it is one of the most demanding 6G service profiles, requiring reliable, low-delay, and low-jitter connectivity~\cite{saad2020vision}. It therefore stresses PLR, delay, and jitter simultaneously, making it suitable for observing the multi-criteria behavior of all compared schemes within one service case. The corresponding VR/AR QoS acceptability ranges are reported in Table~\ref{tab:vrar_qos_ranges}.

\begin{table}[!htbp]
\small
\caption{VR/AR application QoS acceptability ranges used in the evaluation.}
\label{tab:vrar_qos_ranges}
\centering
\vspace{-2mm}
\small
\setlength{\tabcolsep}{2pt}
\begin{tabular}{l|c|c|c}
\hline
\textbf{Criterion} & \textbf{Type} & \textbf{Excellent} & \textbf{Acceptable} \\
\hline
PLR (\%)         & Cost    & 0.10  & 0.30 \\
Delay $D$ (ms)   & Cost    & 1.75  & 2.25 \\
Jitter $J$ (ms)  & Cost    & 0.62  & 0.95 \\
EE (bit/J)       & Benefit & $10^{6}$ & $10^{5}$ \\
Access cost score $C$ & Cost & 2.0 & 7.0 \\
\hline
\end{tabular}
\end{table}

\section{Numerical Results and Discussion}
\label{sec:numerical_results}

\subsection{Cross-Layer DT Data Consistency}
\label{subsec:Cross-Layer}

We  examine the consistency of the generated cross-layer DT dataset to verify that the Sionna RT--to--ns-3 synchronization produces meaningful QoS trends for the considered NS criteria.

\begin{figure*}[!htbp]
    \centering

    \begin{minipage}[t]{0.35\textwidth}
        \centering
        \includegraphics[width=\linewidth]{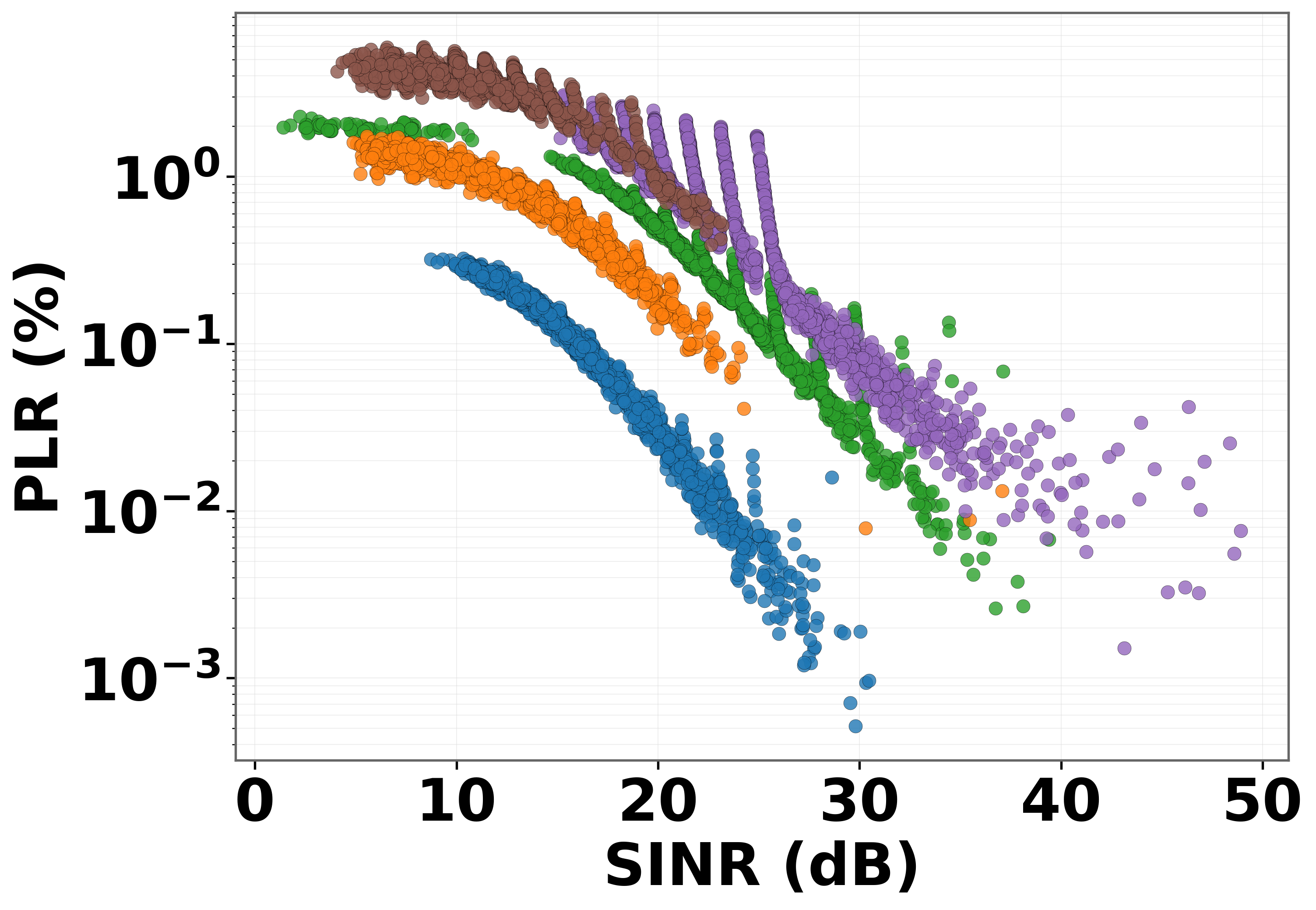}
        \vspace{-6mm}
        \centerline{\small (a) PLR}
    \end{minipage} \hspace{4mm}
    \begin{minipage}[t]{0.35\textwidth}
        \centering
        \includegraphics[width=\linewidth]{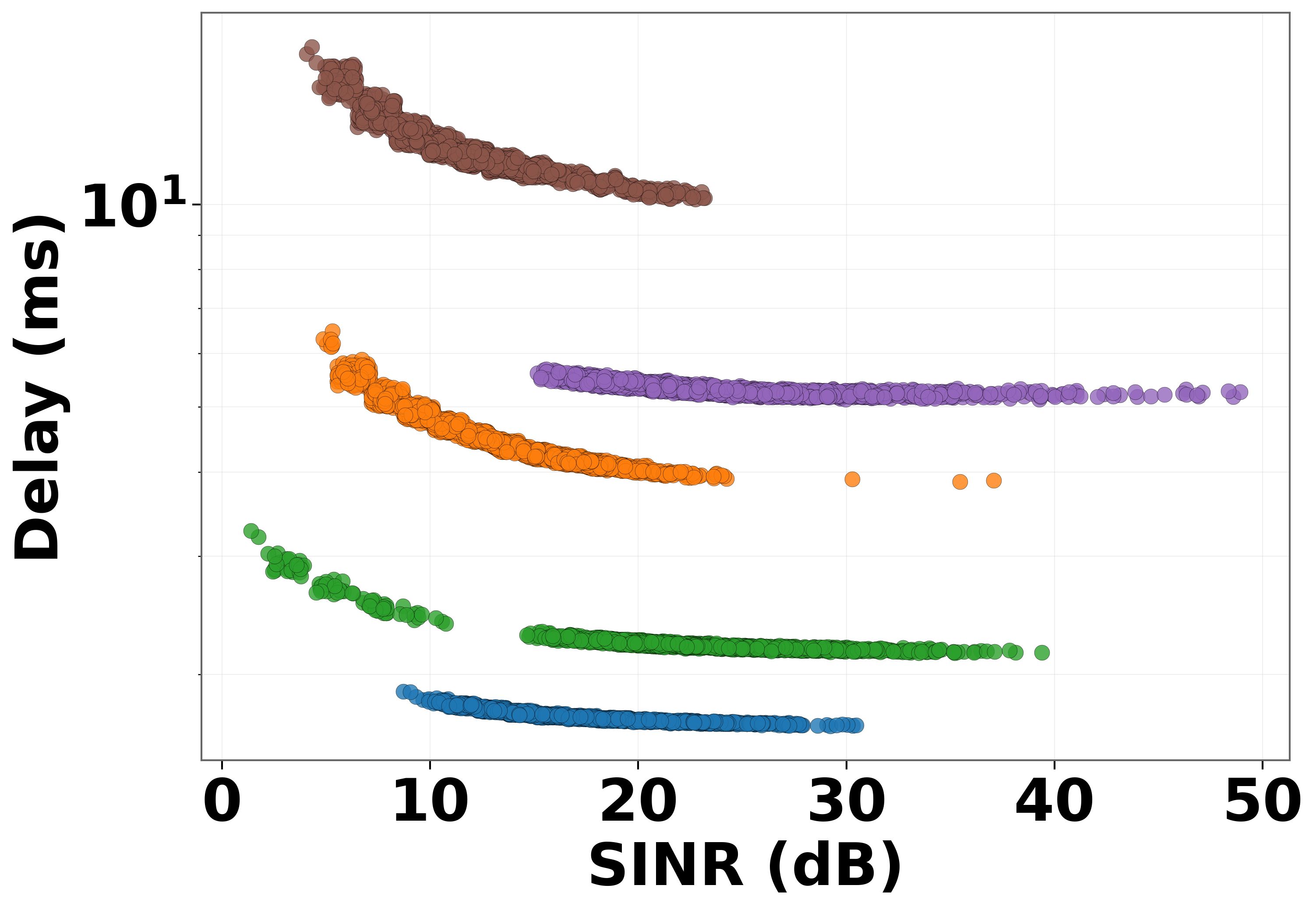}
        \vspace{-2mm}
        \centerline{\small (b) Delay}
    \end{minipage}

    \vspace{4mm}

    \begin{minipage}[t]{0.35\textwidth}
        \centering
        \includegraphics[width=\linewidth]{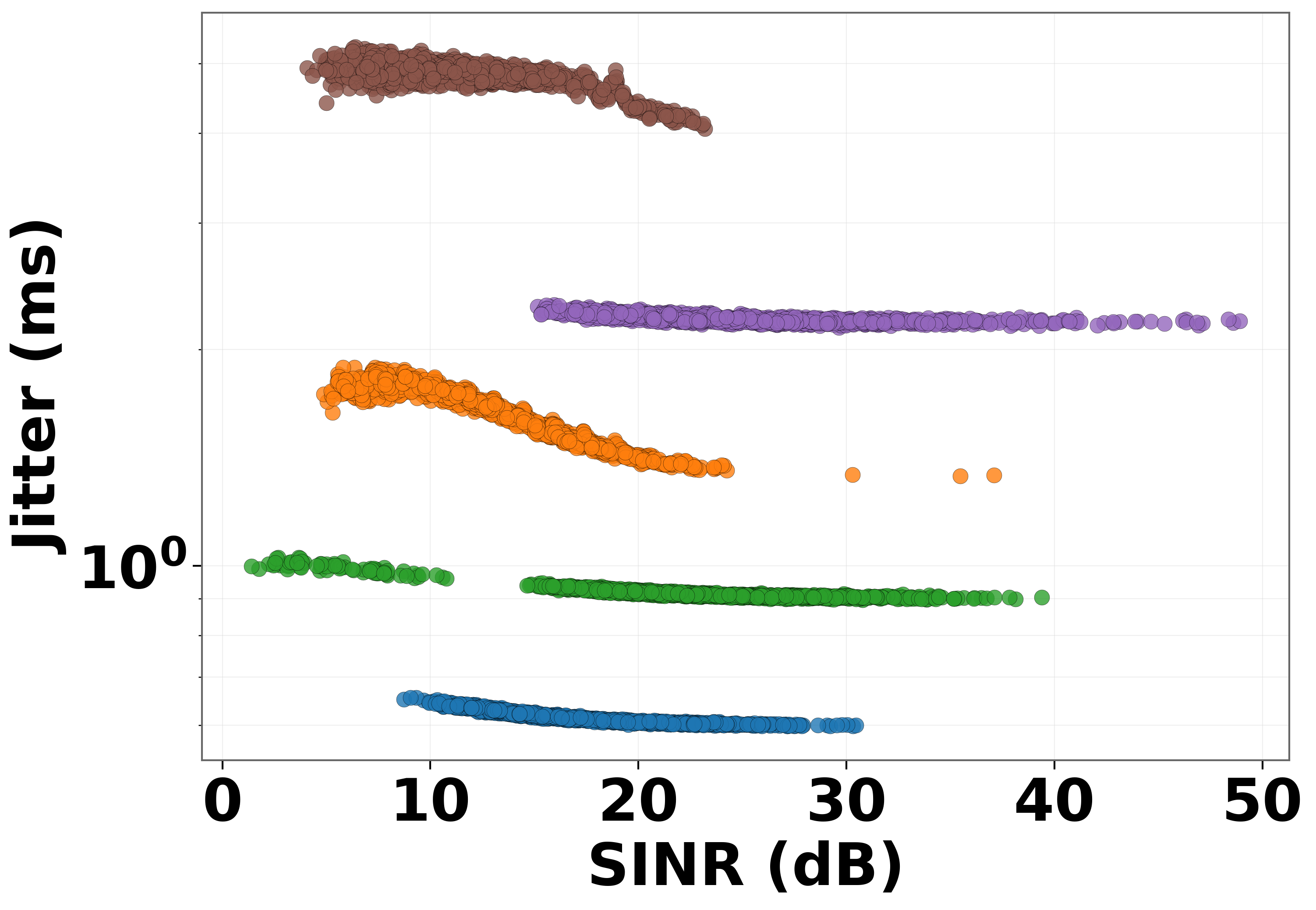}
        \vspace{-4mm}
        \centerline{\small (c) Jitter}
    \end{minipage}\hspace{4mm}
    \begin{minipage}[t]{0.35\textwidth}
        \centering
        \includegraphics[width=\linewidth]{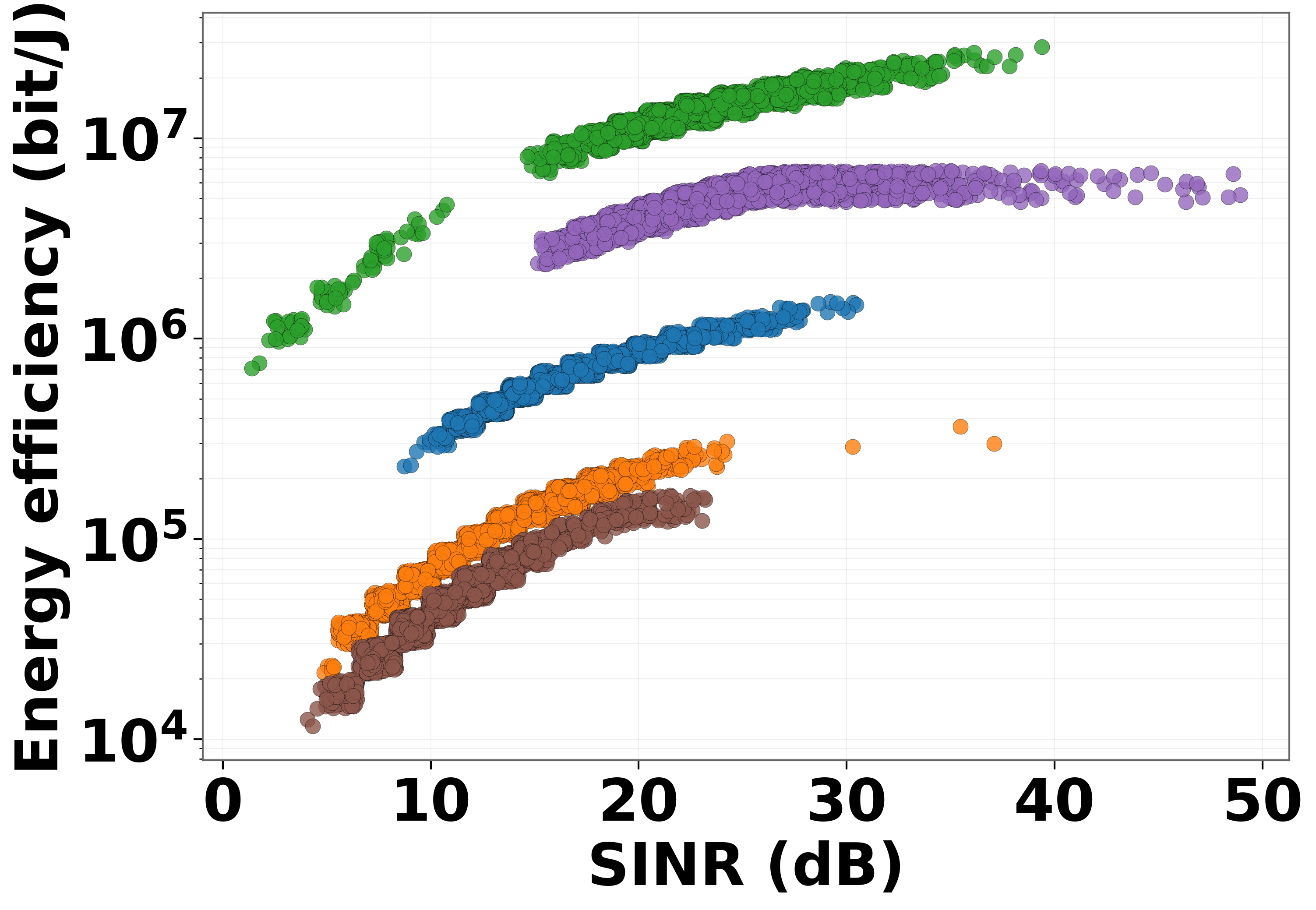}
        \vspace{-2mm}
        \centerline{\small (d) Energy efficiency}
    \end{minipage}

    \vspace{4mm}

    {\small
    \legbox{1F77B4}~5G NR \quad
    \legbox{2CA02C}~Wi-Fi 6 \quad
    \legbox{FF7F0E}~LTE \quad
    \legbox{9467BD}~Wi-Fi \quad
    \legbox{8C564B}~WiMAX
    }

    \caption{Cross-layer DT consistency: derived SINR versus packet-level QoS indicators.}
    \label{fig:dt_consistency}
\end{figure*}

Fig.~\ref{fig:dt_consistency} shows the relationship between the derived SINR and the main NS criteria across the considered decision epochs and candidate RATs. SINR is derived during the Sionna~RT--to--ns-3 synchronization from the geometry-aware channel descriptors and RAT configurations, while PLR, delay, and jitter are obtained from packet-level emulation and EE is computed in post-processing.



As shown in Fig.~\ref{fig:dt_consistency}(a), PLR generally decreases as SINR improves, confirming the expected link-quality behavior. Figs.~\ref{fig:dt_consistency}(b) and~\ref{fig:dt_consistency}(c) show RAT-dependent delay and jitter regions, since these metrics also depend on access timing, queueing, contention, and packet-transmission behavior. Fig.~\ref{fig:dt_consistency}(d) shows that EE improves with link quality but remains strongly RAT-dependent due to differences in bandwidth, achievable rate, and power consumption. The results also highlight the multi-criteria nature of NS. No RAT dominates all criteria simultaneously: 5G NR provides favorable reliability and latency behavior, whereas Wi-Fi~6 achieves substantially higher EE. This motivates application-aware NS, where the preferred RAT depends on the active service profile rather than on a single QoS metric.


Overall, Fig.~\ref{fig:dt_consistency} confirms that the DT-derived dataset preserves both radio-quality effects and RAT-specific packet-level characteristics. Therefore, the subsequent NS performance differences can be attributed to the decision logic rather than to an inconsistent or degenerate QoS trace.

\subsection{Rank Reversal Problem Robustness}
\label{subsec:RRP}

\begin{figure}[!htbp]
    \centering
    
    \begin{minipage}{0.49\textwidth}
        \centering
        \includegraphics[width=0.98\linewidth]{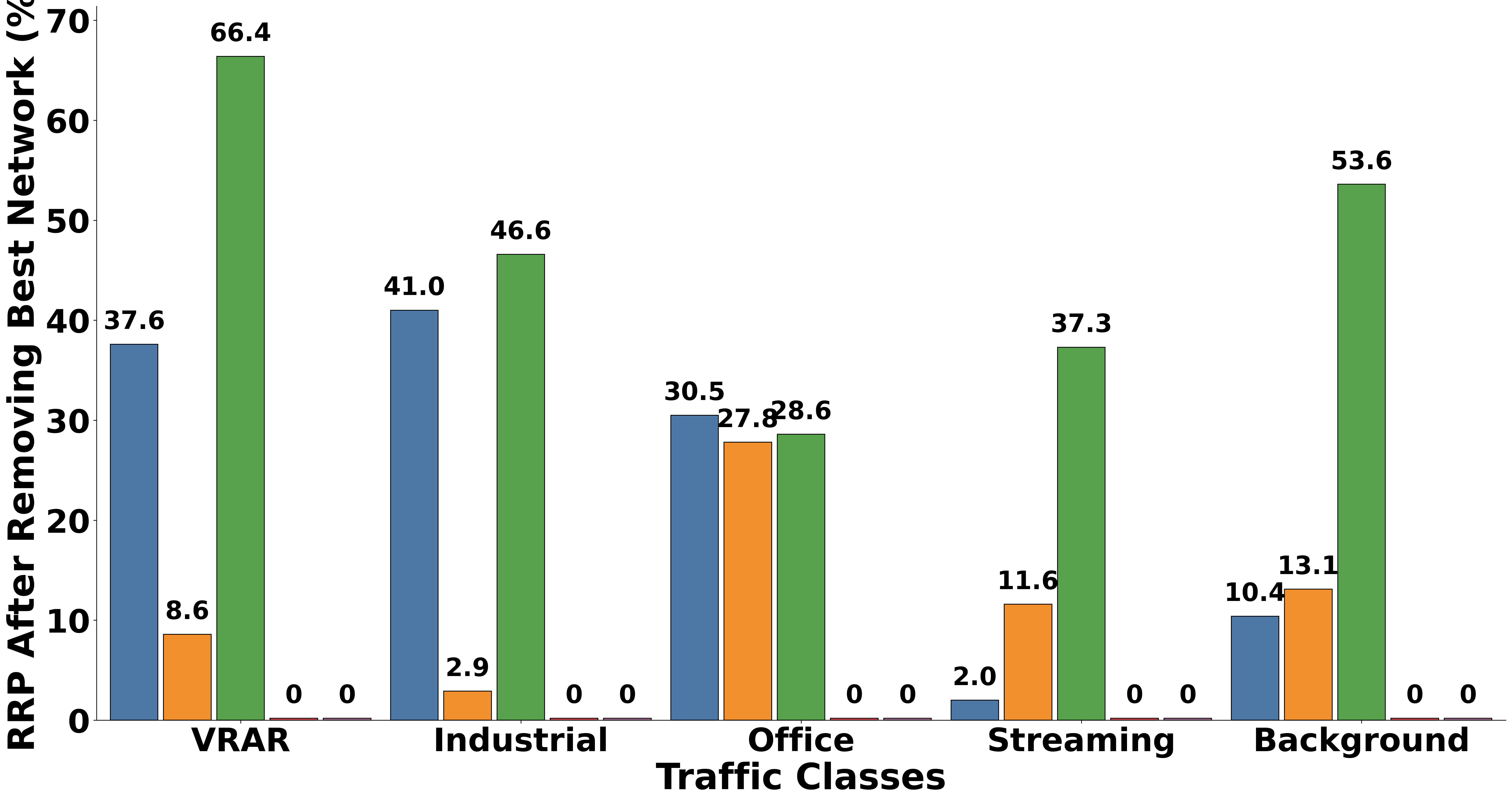}\\
        \vspace{-1mm}
        {\small (a) Remove best-ranked RAT}
    \end{minipage}

    \vspace{0mm}

    \begin{minipage}{0.49\textwidth}
        \centering
        \includegraphics[width=0.98\linewidth]{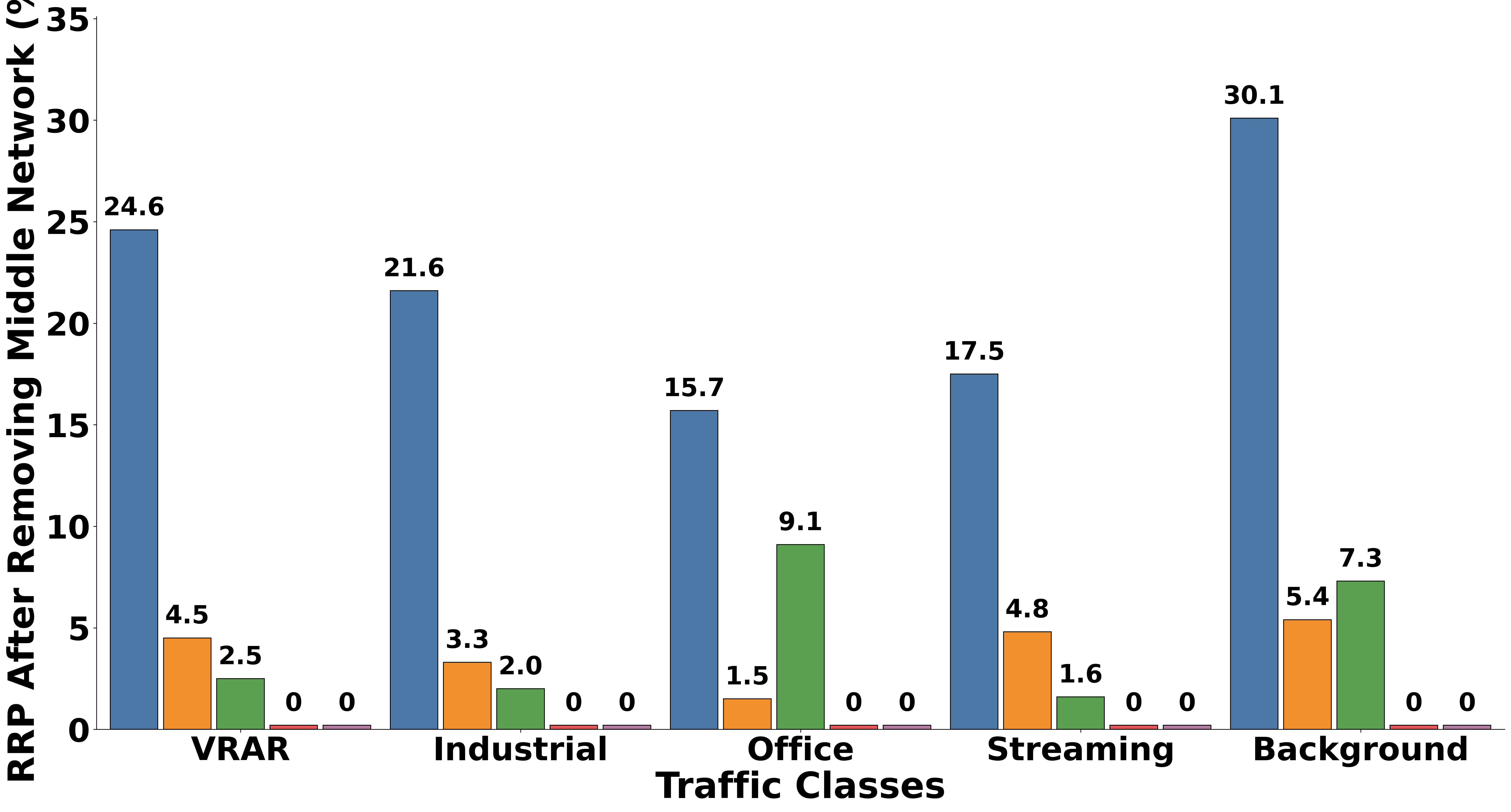}\\
        \vspace{1mm}
        {\small (b) Remove middle-ranked RAT}
    \end{minipage}

    \vspace{0mm}

    \begin{minipage}{0.49\textwidth}
        \centering
        \includegraphics[width=0.98\linewidth]{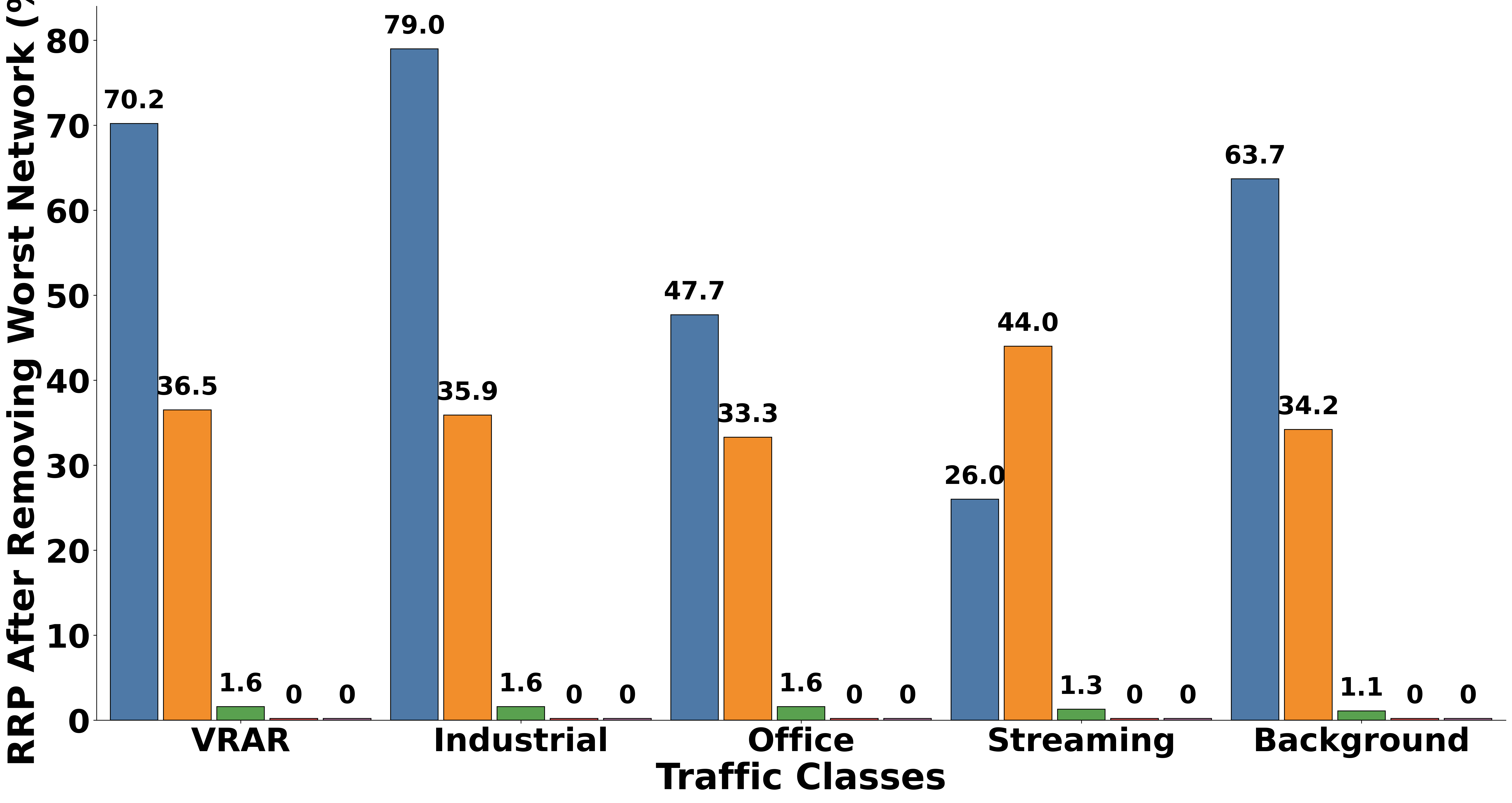}\\
        \vspace{1mm}
        {\small (c) Remove worst-ranked RAT}
    \end{minipage}

    \vspace{0mm}

    \begin{minipage}{0.49\textwidth}
        \centering
        \includegraphics[width=0.98\linewidth]{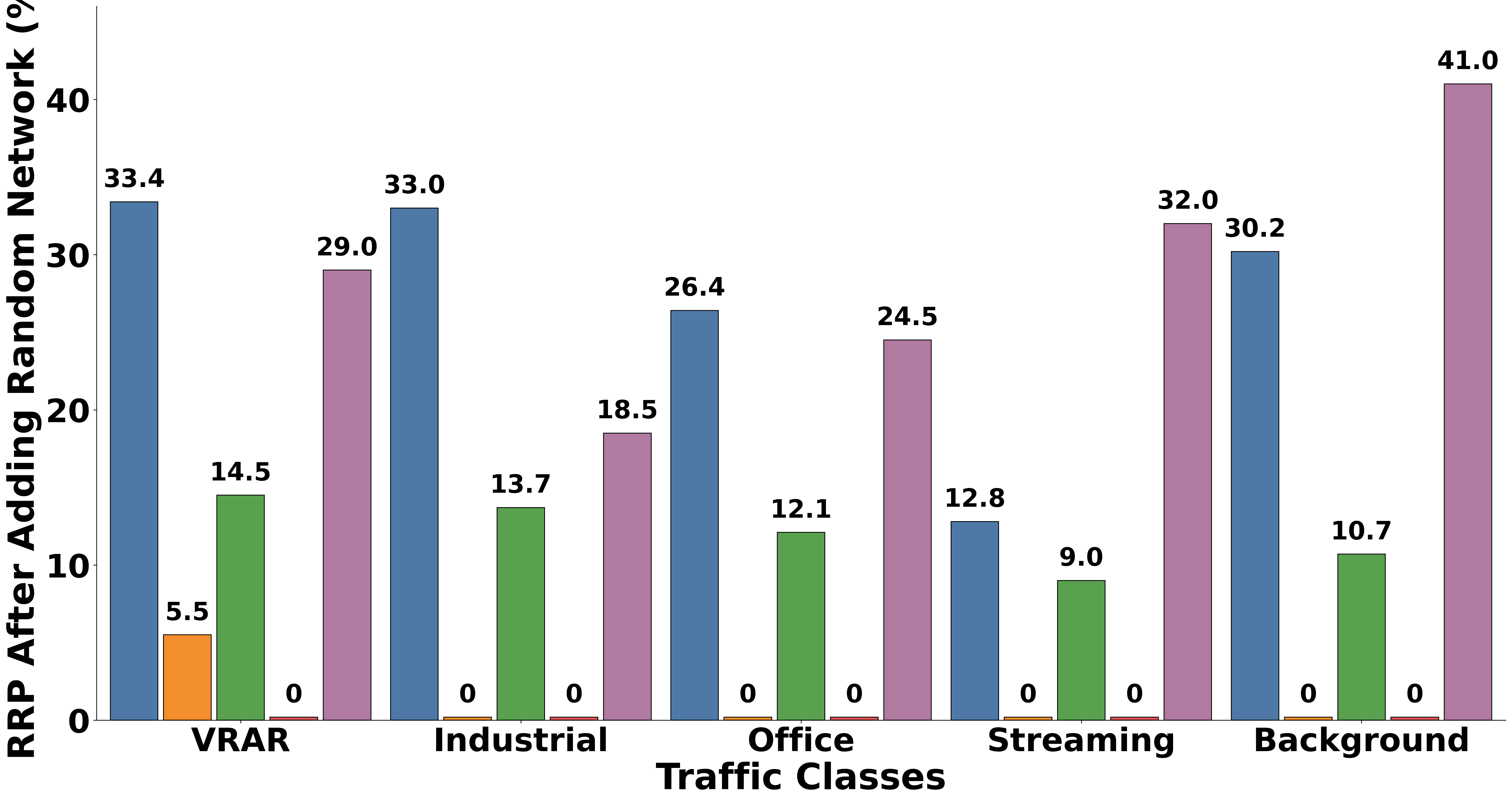}\\
        \vspace{1mm}
        {\small (d) Add a random RAT}
    \end{minipage}

    \vspace{2mm}

    {\small
    \legbox{4E79A7} TOPSIS--AHP \quad
    \legbox{F28E2B} TOPSIS--LLM--NS \quad
    \legbox{59A14F} SAW--AHP \quad
    \legbox{E15759} SAW--LLM--NS \quad
    \legbox{B07AA1} LLM--NS
    }

    \vspace{2mm}

    \caption{RRP under DT-state perturbations.}
    \label{fig:rrp_all}
\end{figure}

Fig.~\ref{fig:rrp_all} reports the RRP rate of each scheme under the four DT-state perturbations of the candidate set. The AHP baselines exhibit strong instability. TOPSIS--AHP is particularly sensitive to worst-removal, reaching up to $79.0\%$ RRP for the industrial profile, while SAW--AHP degrades under best-removal, reaching $66.4\%$ for VR/AR. Both also exhibit noticeable instability under candidate insertion, confirming that fixed AHP weighting over instantaneous decision matrices cannot guarantee stable RAT ordering.

TOPSIS--LLM--NS significantly reduces these reversals by using HAAN with DT-memory-based normalization references. For example, the VR/AR middle-removal RRP decreases from $24.6\%$ to $4.5\%$, while the add-RAT case remains below $1\%$ for most service profiles. The remaining reversals are mainly caused by the deterministic TOPSIS re-evaluation of ideal and anti-ideal points after perturbation.  In contrast, SAW--LLM--NS achieves $0\%$ RRP across all perturbations because HAAN-stabilized linear aggregation preserves the relative scores of retained candidates.


LLM--NS is structurally invariant to candidate withdrawal due to the projection in~\eqref{eq:rag_remove}, resulting in $0\%$ RRP for all removal cases. Under insertion, the LLM evaluates the newly observed RAT within the enlarged candidate context, and residual reversals may occur. The insertion RRP values between $18.5\%$ and $41.0\%$ therefore reflect the stability--adaptivity trade-off discussed in Section~\ref{subsec:rag_icl}: the method preserves retained-candidate order under withdrawal, while allowing adaptation when new RAT evidence is introduced. These insertion-time reversals remain comparable to or below the AHP baselines. Overall, SAW--LLM--NS provides the strongest robustness, followed by TOPSIS--LLM--NS, confirming the importance of DT-memory-assisted normalization and ranking stabilization under DT-state evolution.

\subsection{Achieved QoS Satisfaction}

\begin{figure*}[!htbp]
    \centering

    \begin{minipage}[t]{0.33\textwidth}
        \centering
        \includegraphics[width=0.9\linewidth]{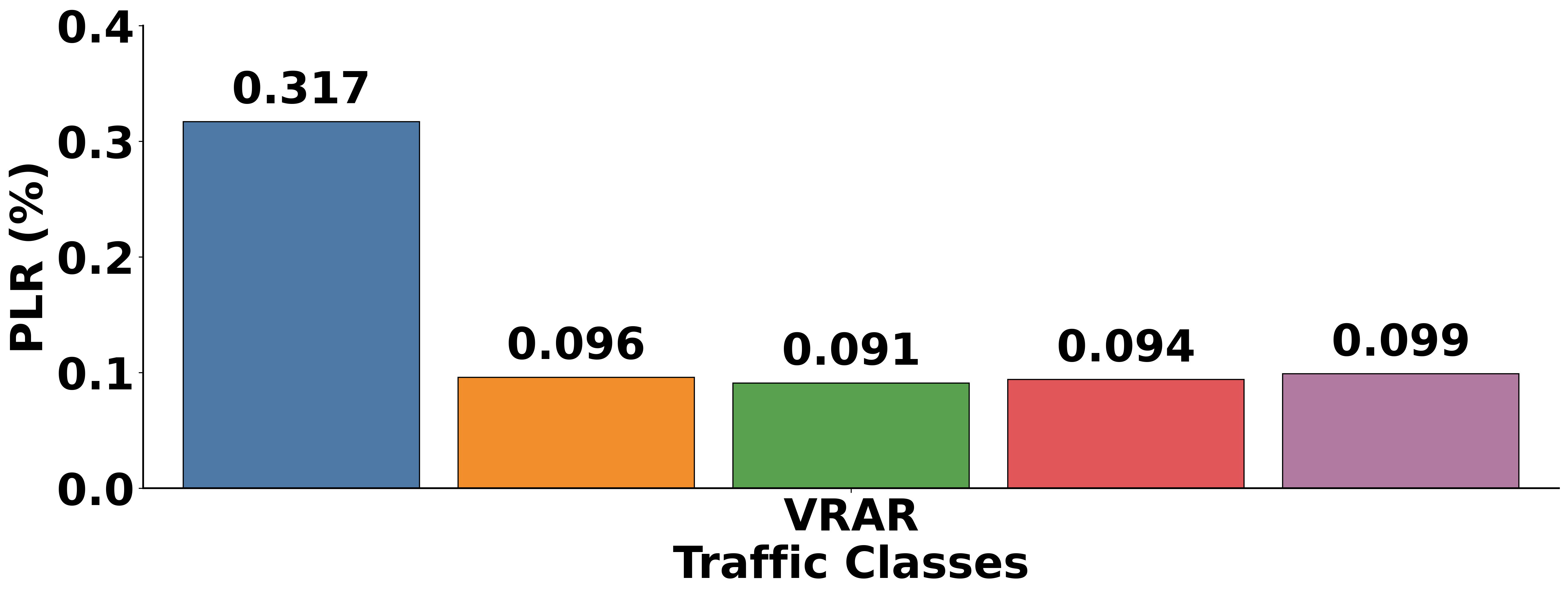}
        \vspace{-2mm}
        \centerline{\small (a) PLR}
    \end{minipage}\hfill
    \begin{minipage}[t]{0.33\textwidth}
        \centering
        \includegraphics[width=0.9\linewidth]{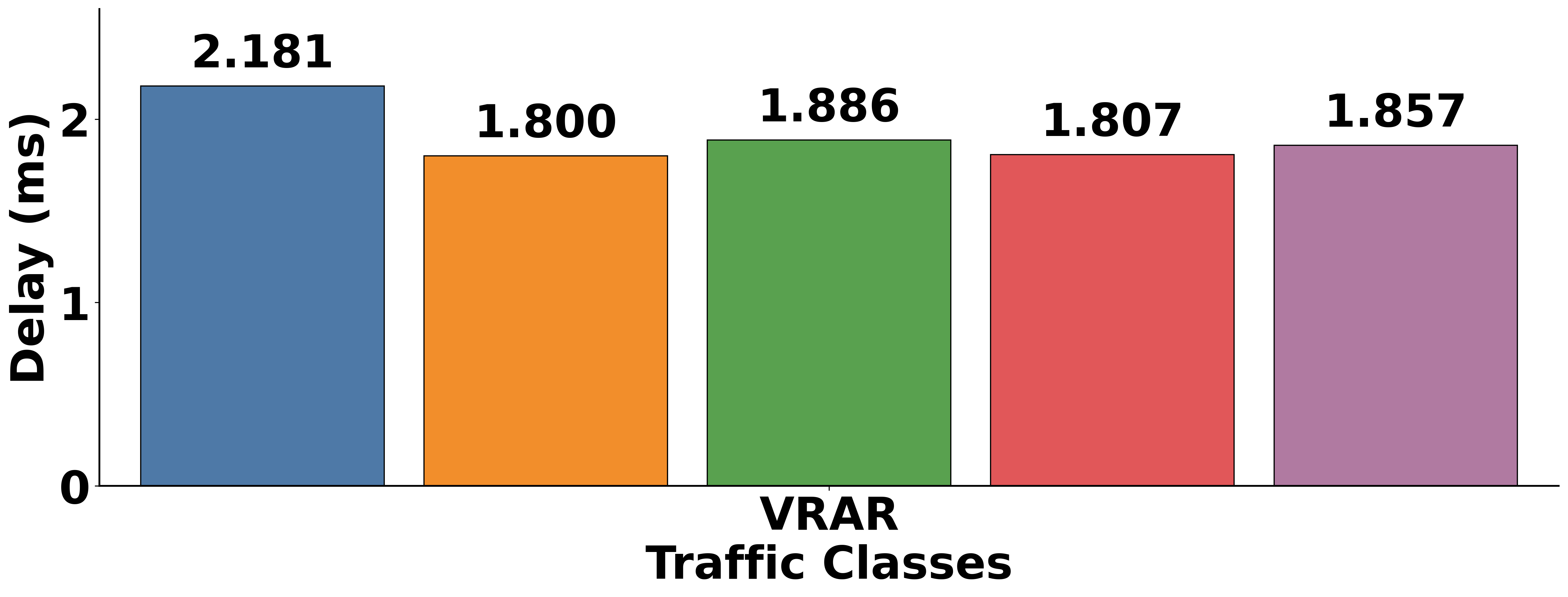}
        \vspace{-2mm}
        \centerline{\small (b) Delay}
    \end{minipage}\hfill
    \begin{minipage}[t]{0.33\textwidth}
        \centering
        \includegraphics[width=0.9\linewidth]{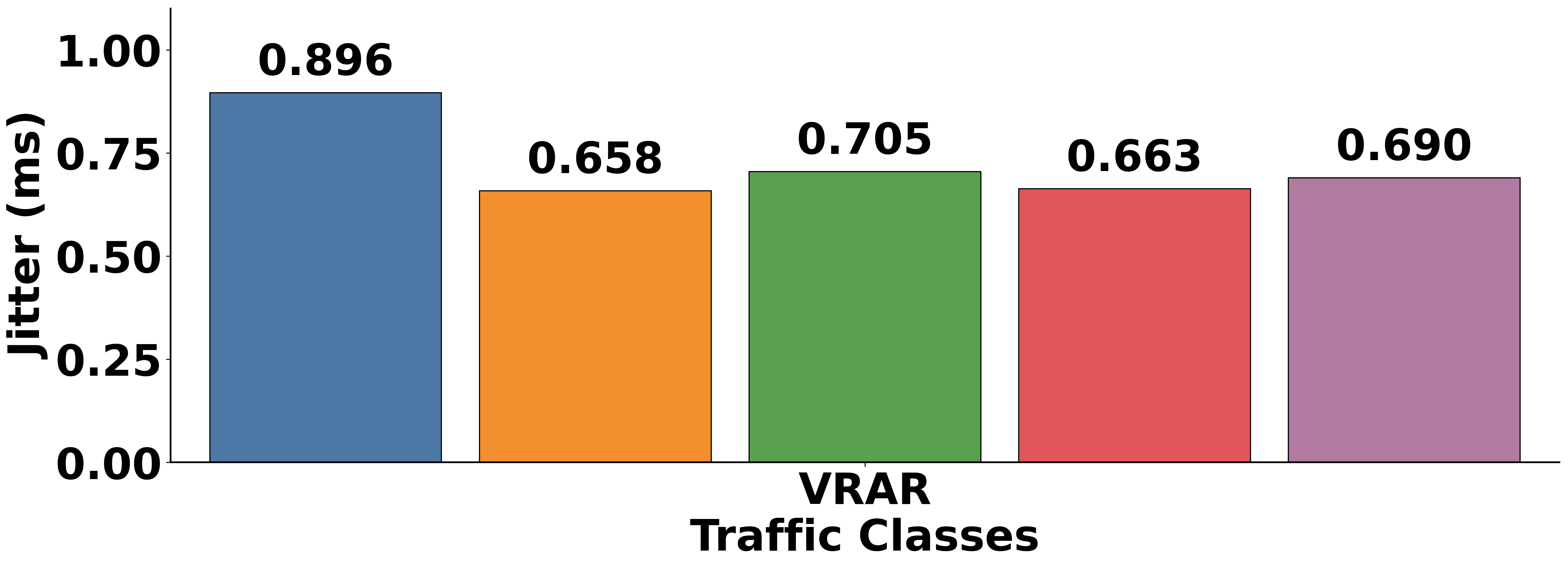}
        \vspace{-2mm}
        \centerline{\small (c) Jitter}
    \end{minipage}

    \vspace{4mm}

\begin{minipage}[c]{0.32\textwidth}
    \centering
    {\small
    \begin{tabular}{@{}l@{\hspace{8pt}}l@{}}
        \legbox{4E79A7}~TOPSIS--AHP  & \legbox{F28E2B}~TOPSIS--LLM--NS \\[2pt]
        \legbox{59A14F}~SAW--AHP     & \legbox{E15759}~SAW--LLM--NS    \\[2pt]
        \legbox{B07AA1}~LLM--NS      &                                  \\
    \end{tabular}}
\end{minipage}\hfill
    \begin{minipage}[c]{0.33\textwidth}
        \centering
        \includegraphics[width=0.9\linewidth]{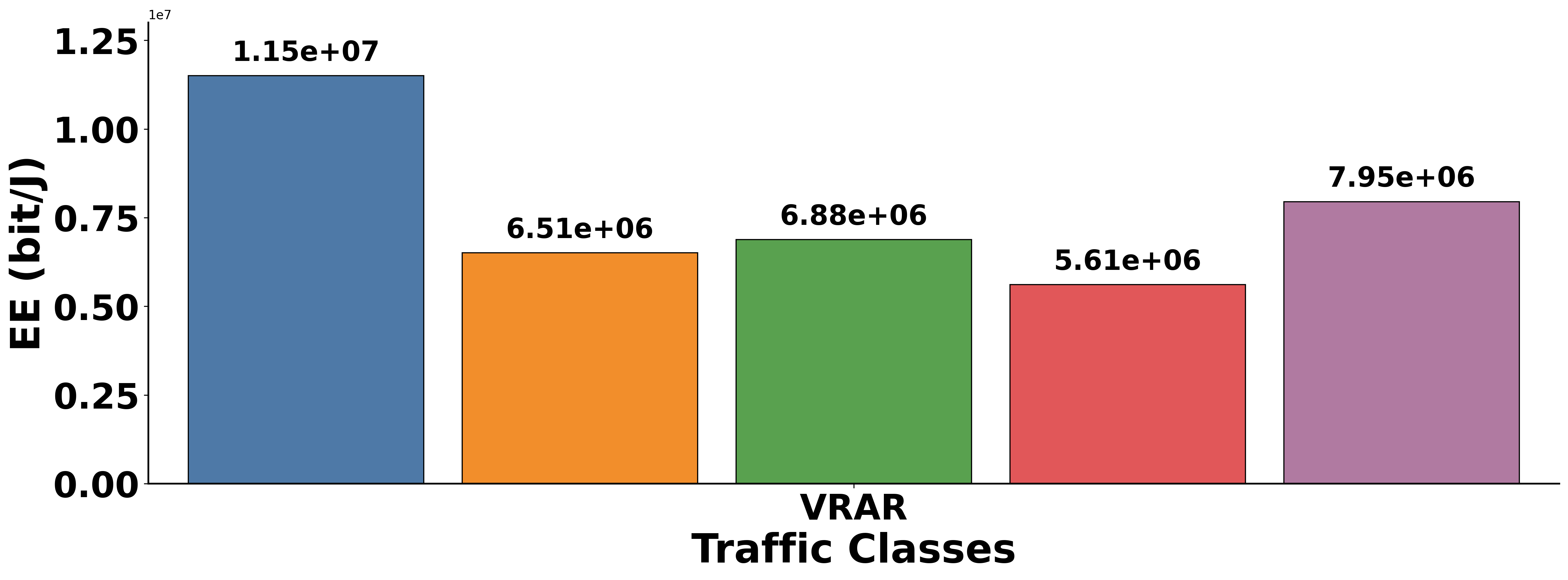}
        \vspace{-2mm}
        \centerline{\small (d) Energy efficiency}
    \end{minipage}\hfill
    \begin{minipage}[c]{0.33\textwidth}
        \centering
        \includegraphics[width=0.9\linewidth]{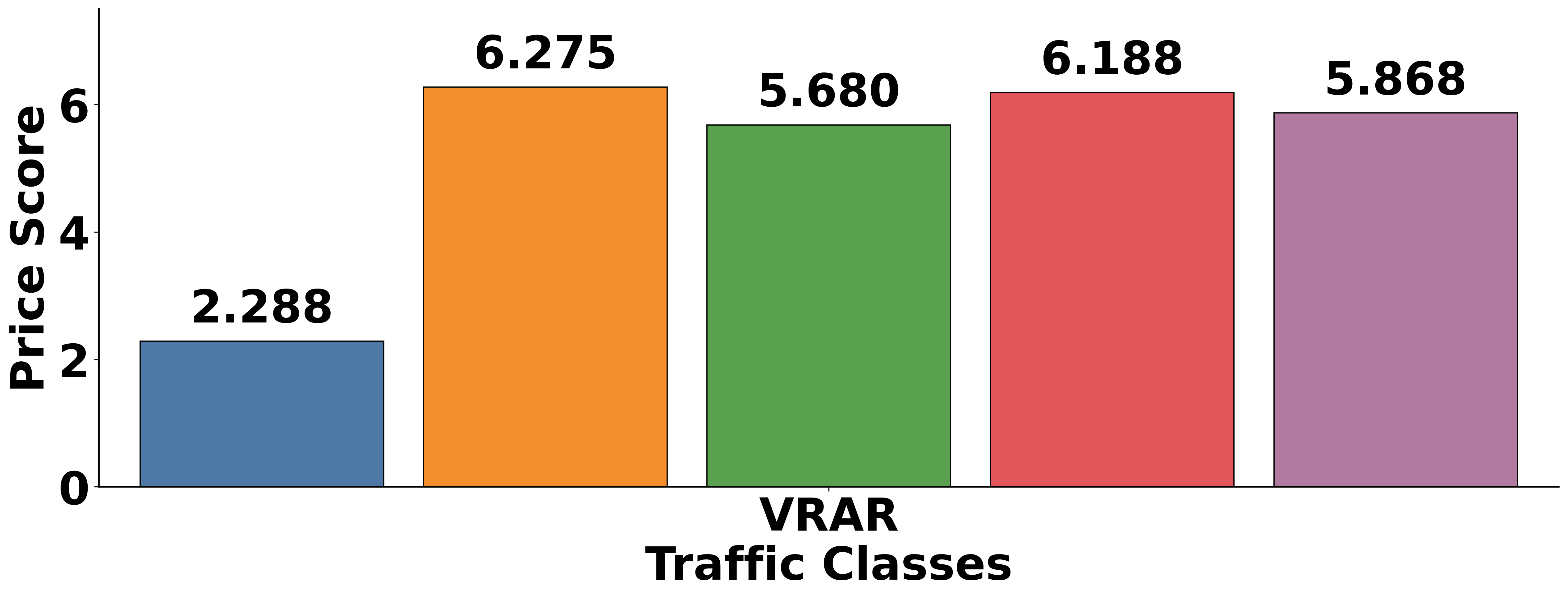}
        \vspace{-2mm}
        \centerline{\small (e) Access cost score}
    \end{minipage}

    \vspace{2mm}

    \caption{QoS performance of the selected RATs under the VR/AR service profile. Subfigures (a)--(e) report PLR, mean delay, mean absolute jitter, EE, and access-cost score, respectively.}
    \label{fig:qos_vrar_results}
\end{figure*}

Fig.~\ref{fig:qos_vrar_results} reports the KPI-level performance of the RATs selected by each scheme under the VR/AR service profile. Since VR/AR is highly sensitive to reliability and latency, the main comparison focuses on PLR, delay, and jitter, while EE and access cost are used to interpret the resulting trade-offs.

The results show that the LLM-based schemes improve reliability-oriented RAT selection. In particular, LLM--NS achieves the lowest PLR at $0.26\%$, compared with $0.41\%$ for TOPSIS--AHP and $0.46\%$ for SAW--AHP. The MADM--LLM--NS schemes also reduce PLR to $0.29\%$, showing that LLM-driven intent interpretation improves the reliability alignment of deterministic MADM ranking. For delay and jitter, TOPSIS--LLM--NS maintains low delay while reducing jitter to $0.02$ ms, whereas LLM--NS provides the best overall balance across PLR, delay, and jitter.

The EE and access-cost results further highlight the multi-criteria trade-off. TOPSIS--AHP achieves the highest EE and lowest cost, but with higher PLR. In contrast, LLM--NS maintains competitive EE and moderate cost while improving the primary VR/AR reliability metric. Overall, the proposed LLM-based schemes provide more service-aware RAT selection, mainly through better alignment with the dominant VR/AR QoS requirements.

\subsection{Unnecessary Handover}
\label{subsec:unnecessary_handover}

\begin{table}[!htbp]
\small
\caption{Unnecessary VHO ratio and reduction under the VR/AR profile.}
\label{tab:uho_summary}
\centering
\setlength{\tabcolsep}{5pt}
\begin{tabular}{l|p{2cm}|c}
\hline
\textbf{Scheme} & \textbf{Unnec. VHO (\%)} & \textbf{Reduction vs. AHP (\%)} \\
\hline
TOPSIS--AHP & 7.111 & --- \\
TOPSIS--LLM--NS & 5.111 & 28.1 \\
SAW--AHP & 10.667 & --- \\
SAW--LLM--NS & 5.556 & 47.9 \\
LLM--NS & 7.556 & --- \\
\hline
\end{tabular}
\end{table}

Table~\ref{tab:uho_summary} reports the unnecessary VHO ratio under the VR/AR service profile. A lower value indicates fewer RAT changes that do not provide sufficient service-level improvement. The reduction column is computed relative to the corresponding AHP baseline for each MADM operator. TOPSIS--LLM--NS reduces the unnecessary VHO ratio from $7.11\%$ to $5.11\%$, corresponding to a $28.1\%$ reduction compared with TOPSIS--AHP. Similarly, SAW--LLM--NS reduces the unnecessary VHO ratio from $10.66\%$ to $5.55\%$, corresponding to a $47.9\%$ reduction compared with SAW--AHP.

These results show that the proposed MADM--LLM--NS schemes improve mobility stability by reducing RAT changes that do not yield sufficient service-level benefit. This improvement is mainly due to the combination of UIA-driven service-aware weighting and HAAN-based DT-memory normalization, which makes the MADM ranking less sensitive to instantaneous candidate-set fluctuations. LLM--NS achieves an unnecessary VHO ratio of $7.55\%$. This value is higher than those of the HAAN-stabilized MADM--LLM--NS branches, reflecting the more adaptive behavior of direct listwise LLM ranking when newly observed candidate-RAT evidence is introduced. Overall, the results indicate a stability--adaptivity trade-off: MADM--LLM--NS provides the strongest unnecessary-VHO reduction, while LLM--NS remains more adaptive to candidate-set evolution.

\subsection{Computational Complexity}
\label{subsec:complexity_analysis}

We analyze the computational complexity of the proposed DT-grounded framework by separating scenario-level DT generation from per-epoch decision-layer processing. Let \(n\) denote the number of candidate RATs, \(m\) the number of decision criteria, \(K\) the HAAN memory-window size, and \(\mathcal{M}_k\) the DT memory at epoch \(k\). As summarized in Table~\ref{tab:complexity_overhead}, scene construction, geometry-aware ray tracing, and  packet-level emulation are scenario-level DT-generation costs, denoted by \(C_{\mathrm{scene}}+C_{\mathrm{RT}}+C_{\mathrm{ns3}}\). Once the DT-derived candidate state is available, forming the decision state requires \(O(nm)\) per epoch.

\begin{table}[!htbp]
\caption{Computational complexity of the proposed framework.}
\label{tab:complexity_overhead}
\centering
\setlength{\tabcolsep}{4pt}
\begin{tabularx}{\columnwidth}{l X}
\hline
\textbf{Component / branch} & \textbf{Complexity} \\
\hline
Scenario-level DT construction & $C_{\mathrm{scene}}+C_{\mathrm{RT}}+C_{\mathrm{ns3}}$ \\
DT decision-state formation & $O(nm)$ \\
TOPSIS--AHP, SAW--AHP & $O(nm+n\log n)$ \\
TOPSIS--LLM--NS, SAW--LLM--NS & $O(C_{\mathrm{LLM}}^{\mathrm{UIA}}+nm+Km+n\log n)$ \\
LLM--NS, memory hit & $O(|\mathcal{M}_k|\log|\mathcal{M}_k|+n\log n)$ \\
LLM--NS, memory miss & $O(|\mathcal{M}_k|\log|\mathcal{M}_k|+C_{\mathrm{LLM}}^{\mathrm{rank}}+nm+n\log n)$ \\
Evaluation and memory update & $O(m)$ \\
\hline
\end{tabularx}
\end{table}

For the decision branches, TOPSIS--AHP and SAW--AHP require \(O(nm+n\log n)\), accounting for criterion processing, score computation, and candidate ranking. TOPSIS--LLM--NS and SAW--LLM--NS add the UIA inference cost \(C_{\mathrm{LLM}}^{\mathrm{UIA}}\), since the service-aware weights are generated from the current DT-derived decision context, and they also include the HAAN update \(O(Km)\). Their complexity is therefore \(O(C_{\mathrm{LLM}}^{\mathrm{UIA}}+nm+Km+n\log n)\). For LLM--NS, the DT memory acts as a reuse mechanism rather than merely an additional prompt source. If a relevant previous DT state is retrieved and its stored ranking can be reused, the LLM ranking call is avoided and the cost reduces to retrieval and ordering. Otherwise, the branch performs direct listwise LLM ranking, giving \(O(|\mathcal{M}_k|\log|\mathcal{M}_k|+C_{\mathrm{LLM}}^{\mathrm{rank}}+nm+n\log n)\). Thus, the main conclusion is that MADM--LLM--NS pays a lighter LLM call for intent-to-weight generation followed by deterministic MADM ranking, whereas LLM--NS becomes expensive only on memory misses that require a fresh listwise LLM decision. This makes DT-memory reuse central to reducing the per-epoch cost of the direct LLM--NS branch.



\section{Conclusion}
\label{sec:conclusion}

This paper proposed a DT-grounded LLM intelligence framework for application-aware NS in 6G HWNs. The framework shifts NS from an instantaneous decision-matrix operation to a decision process conditioned on an evolving cross-layer DT state, including geometry-aware channel descriptors, packet-level QoS indicators, service intent, candidate-RAT context, and DT memory. Within this framework, the UIA translates service/user intent into structured criterion priorities, while the two decision branches implement LLM-assisted MADM ranking with HAAN and direct LLM-based listwise ranking with RA--ICL.

The numerical results show that memory-assisted DT reasoning improves decision stability under candidate-set evolution. Across the five service profiles, HAAN substantially reduces normalization-induced RRP in the MADM--LLM--NS branch, while RA--ICL preserves retained-candidate ordering under withdrawal and allows insertion-time adaptation when newly observed RAT evidence is introduced. For the VR/AR profile, the LLM-based schemes improve reliability-oriented RAT selection by reducing PLR while maintaining favorable delay and jitter behavior. The unnecessary-VHO analysis further shows that HAAN-stabilized MADM--LLM--NS reduces unstable RAT changes that do not provide sufficient service-level benefit.

Overall, the results support the central idea of this work: application-aware NS should not rely only on isolated instantaneous decision matrices, but should operate over an evolving DT state that combines cross-layer evidence, service intent, and decision memory. Future work will investigate agentic and multi-agent LLM architectures for autonomous DT operation, compare the impact of different LLM families and model scales on NS reliability and latency, and extend the framework toward multimodal DT intelligence by incorporating radio, traffic, map, and visual-context information.

\balance

\bibliographystyle{IEEEtran}

\bibliography{Bibliography}

@article{trestian2012game,
  title={Game theory-based network selection: Solutions and challenges},
  author={Trestian, Ramona and Ormond, Olga and Muntean, Gabriel-Miro},
  journal={IEEE Commun. Surveys Tuts.},
  volume={14},
  number={4},
  pages={1212--1231},
  year={2012},
  publisher={IEEE}
}

@article{van2025network,
  title={Network Access Selection for {URLLC} and {eMBB} Applications in Sub-{6GHz}-mmWave-{THz} Networks: Game Theory Versus Multi-Agent Reinforcement Learning},
  author={Van, Nguyen Thi Thanh and others},
  journal={IEEE Trans. Commun.},
  year={2025},
  publisher={IEEE}
}

@ARTICLE{9351659,
  author={Chkirbene, Zina and Abdellatif, Alaa Awad and Mohamed, Amr and Erbad, Aiman and Guizani, Mohsen},
  journal={IEEE Transactions on Network Science and Engineering}, 
  title={Deep Reinforcement Learning for Network Selection Over Heterogeneous Health Systems}, 
  year={2022},
  volume={9},
  number={1},
  pages={258-270},
  doi={10.1109/TNSE.2021.3058037}}

@article{allahham2022multi,
  title={Multi-agent reinforcement learning for network selection and resource allocation in heterogeneous multi-RAT networks},
  author={Allahham, Mhd Saria and Abdellatif, Alaa Awad and Mhaisen, Naram and Mohamed, Amr and Erbad, Aiman and Guizani, Mohsen},
  journal={IEEE Transactions on Cognitive Communications and Networking},
  volume={8},
  number={2},
  pages={1287--1300},
  year={2022},
  publisher={IEEE}
}

@article{mefgouda2025balancing,
  title={Balancing Subjectivity and Objectivity in Network Selection: A Decision-Making Framework Towards Digital Twins},
  author={Mefgouda, Brahim and Idoudi, Hanen and Al-Quraan, Mohammad and Lotfi, Ismail and Alhussein, Omar and Mohjazi, Lina and Muhaidat, Sami},
  journal={arXiv preprint arXiv:2504.01414},
  year={2025}
}

@article{wang2012mathematical,
  title={Mathematical modeling for network selection in heterogeneous wireless networks—A tutorial},
  author={Wang, Lusheng and Kuo, Geng-Sheng GS},
  journal={IEEE Commun. Surveys Tuts.},
  volume={15},
  number={1},
  pages={271--292},
  year={2012},
  publisher={IEEE}
}

@article{obayiuwana2017network,
  title={Network selection in heterogeneous wireless networks using multi-criteria decision-making algorithms: a review},
  author={Obayiuwana, Enoruwa and Falowo, Olabisi Emmanuel},
  journal={Wireless Networks},
  volume={23},
  number={8},
  pages={2617--2649},
  year={2017},
  publisher={Springer}
}

@article{senouci2016topsis,
  title={TOPSIS-based dynamic approach for mobile network interface selection},
  author={Senouci, Mohamed Abdelkrim and Mushtaq, M Sajid and Hoceini, Said and Mellouk, Abdelhamid},
  journal={Computer Networks},
  volume={107},
  pages={304--314},
  year={2016},
  publisher={Elsevier}
}

@inproceedings{mefgouda2021cocoso,
  title={COCOSO-based network interface selection algorithm for heterogeneous wireless networks},
  author={Mefgouda, Brahim and Idoudi, Hanen},
  booktitle={2021 International Conference on Innovation and Intelligence for Informatics, Computing, and Technologies (3ICT)},
  pages={1--5},
  year={2021},
  organization={IEEE}
}

@article{dos2025network,
  title={Network Selection in B5G Heterogeneous Networks Using AHP-TOPSIS},
  author={Dos Santos, Albert and Pantoja, Matheus and Paix{\~a}o, Erm{\'\i}nio and Teixeira, Reyso and Seruffo, Marcos and Cardoso, Diego},
  journal={IEEE Access},
  year={2025},
  publisher={IEEE}
}

@article{ndashimye2021multi,
  title={A Multi-criteria based handover algorithm for vehicle-to-infrastructure communications},
  author={Ndashimye, Emmanuel and Sarkar, Nurul I and Ray, Sayan Kumar},
  journal={Computer Networks},
  volume={185},
  pages={107652},
  year={2021},
  publisher={Elsevier}
}

@article{vaidya2006analytic,
  title={Analytic hierarchy process: An overview of applications},
  author={Vaidya, Omkarprasad S and Kumar, Sushil},
  journal={European Journal of operational research},
  volume={169},
  number={1},
  pages={1--29},
  year={2006},
  publisher={Elsevier}
}

@article{jahan2015state,
  title={A state-of-the-art survey on the influence of normalization techniques in ranking: Improving the materials selection process in engineering design},
  author={Jahan, Ali and Edwards, Kevin L},
  journal={Mater. Des.},
  volume={65},
  pages={335--342},
  year={2015},
  publisher={Elsevier}
}

@article{tu2025analytic,
  title={Analytic hierarchy process rank reversals: causes and solutions},
  author={Tu, Jiancheng and Wu, Zhibin},
  journal={Annals of Operations Research},
  volume={346},
  number={2},
  pages={1785--1809},
  year={2025},
  publisher={Springer}
}

@inproceedings{senouci2016utility,
  title={Utility function-based {TOPSIS} for network interface selection in heterogeneous wireless networks},
  author={Senouci, Mohamed Abdelkrim and Hoceini, Said and Mellouk, Abdelhamid},
  booktitle={2016 IEEE international conference on communications (ICC)},
  pages={1--6},
  year={2016},
month=may,
address={Kuala Lumpur, Malaysia},
  organization={IEEE}
}

@article{silva2024comprehensive,
  title={A comprehensive step-wise survey of multiple attribute decision-making mobility approaches},
  author={Silva, Felipe S Dantas and Lima, Mathews PS and Corujo, Daniel and Neto, Augusto J Ven{\^a}ncio and Esposito, Flavio},
  journal={IEEE Access},
  volume={12},
  pages={108616--108656},
  year={2024},
  publisher={IEEE}
}

@article{chandavarkar2016simplified,
  title={Simplified and improved multiple attributes alternate ranking method for vertical handover decision in heterogeneous wireless networks},
  author={Chandavarkar, Beerappa R and Guddeti, Ram Mohana Reddy},
  journal={Computer Communications},
  volume={83},
month=jun,
  pages={81--97},
  year={2016},
  publisher={Elsevier}
}

@article{baghla2018vikor,
  title={VIKOR MADM based optimization method for vertical handover in heterogeneous networks},
  author={Baghla, Silki and Bansal, Savina},
  journal={Advances in Systems Science and Applications},
  volume={18},
  number={3},
  pages={90--110},
  year={2018}
}

@article{mansouri2019new,
  title={New Manhattan distance-based fuzzy {MADM} method for the network selection},
  author={Mansouri, Mouad and Leghris, Cherkaoui},
  journal={IET Commun.},
  volume={13},
  number={13},
  pages={1980--1987},
  year={2019},
  publisher={IET}
}

@article{mefgouda2024qos,
  title={A QoS-aware service-driven network selection for HWNs based on MARCOS and utility functions},
  author={Mefgouda, Brahim and Idoudi, Hanen and Al-Quraan, Mohammad and Waqar, Omer and Zoha, Ahmed and Imran, Muhammad Ali and Mohjazi, Lina},
  journal={IEEE Open Journal of the Communications Society},
  volume={5},
  pages={3658--3677},
  year={2024},
  publisher={IEEE}
}

@article{almutairi2018genetic,
  title={A genetic algorithm approach for multi-attribute vertical handover decision making in wireless networks},
  author={Almutairi, Ali F and others},
  journal={Telecommunication Systems},
  volume={68},
  number={2},
  pages={151--161},
month=sep,
  year={2018},
  publisher={Springer}
}

@article{almutairi2021particle,
  title={Particle swarm optimization application for multiple attribute decision making in vertical handover in heterogenous wireless networks},
  author={Almutairi, Ali F and Al-Gharabally, Mishal and Salman, Ayed A},
  journal={J. Eng. Res.},
  volume={9},
  number={1},
  year={2021}
}

@article{mefgouda2023new,
  title={New network interface selection based on {MADM} and multi-objective whale optimization algorithm in heterogeneous wireless networks},
  author={Mefgouda, Brahim and Idoudi, Hanen},
  journal={J. Supercomput.},
  volume={79},
  number={4},
  pages={3580--3615},
  year={2023},
  publisher={Springer}
}

@article{radouche2021new,
  title={New network selection algorithm based on cosine similarity distance and PSO in heterogeneous wireless networks},
  author={Radouche, Said and Leghris, Cherkaoui},
  journal={Journal of Computer Networks and Communications},
  volume={2021},
  number={1},
  pages={6613460},
  year={2021},
  publisher={Wiley Online Library}
}

@article{krishnan2025algorithm,
  title={An algorithm for heterogeneous wireless network connections for user preferences and services},
  author={Krishnan, S Dinesh and Daniel, A and Ayyasamy, S and Balusamy, Balamurugan and Selvarajan, Shitharth and Al-Shehari, Taher and Alsadhan, Nasser A},
  journal={Scientific Reports},
  volume={15},
  number={1},
  pages={17340},
  year={2025},
  publisher={Nature Publishing Group UK London}
}

@article{singh2025intelligent,
  title={Intelligent Network Selection Mechanisms in the Internet of Everything System},
  author={Singh, Indrasen and Munjal, Meenakshi},
  journal={IEEE Access},
  year={2025},
  publisher={IEEE}
}

@article{yu2020novel,
  title={A Novel Heterogeneous Wireless Network Selection Algorithm Based on {INFAHP} and {IGRA}},
  author={YU, HEWEI and GUO, MEIYUAN and YU, JINGXI},
  journal={Journal of Interconnection Networks},
  volume={20},
  number={03},
  year={2020},
  publisher={World Scientific}
}

@article{yu2018heterogeneous,
  title={A heterogeneous network selection algorithm based on network attribute and user preference},
  author={Yu, He-Wei and Zhang, Biao},
  journal={AD hoc Networks},
  volume={72},
month=apr,
  pages={68--80},
  year={2018},
  publisher={Elsevier}
}

@inproceedings{radouche2020network,
  title={Network selection based on cosine similarity and combination of subjective and objective weighting},
  author={Radouche, Said and Leghris, Cherkaoui},
  booktitle={2020 International Conference on Intelligent Systems and Computer Vision (ISCV)},
  pages={1--7},
  year={2020},
  address={Fez, Morocco},
month=jun,
  organization={IEEE}
}

@standard{3gpp_ts_23_107_v18,
  title        = {Quality of Service (QoS) Concept and Architecture},
  organization = {3GPP},
  number       = {TS 23.107},
  edition      = {v18.0.0},
  note         = {Release 18; published as ETSI TS 123 107 V18.0.0},
  year         = {2024},
  month        = apr
}

@incollection{riley2010ns,
  title={The ns-3 network simulator},
  author={Riley, George F and Henderson, Thomas R},
  booktitle={Modeling and tools for network simulation},
  pages={15--34},
  year={2010},
  publisher={Springer}
}

@article{zhu2021adaptive,
  title={Adaptive multi-access algorithm for multi-service edge users in 5G ultra-dense heterogeneous networks},
  author={Zhu, Anqi and Ma, Mingfang and Guo, Songtao and Yu, Shui and Yi, Lin},
  journal={IEEE Transactions on Vehicular Technology},
  volume={70},
  number={3},
  pages={2807--2821},
  year={2021},
  publisher={IEEE}
}

@article{saad2020vision,
  author  = {Saad, Walid and Bennis, Mehdi and Chen, Mingzhe},
  title   = {A Vision of 6G Wireless Systems: Applications, Trends, Technologies, and Open Research Problems},
  journal = {IEEE Network},
  volume  = {34},
  number  = {3},
  pages   = {134--142},
  year    = {2020},
  doi     = {10.1109/MNET.001.1900287}
}

@article{bariah2025waves,
  title={From Waves to Words: Large Perceptive Models as the Future of IoT and RF Intelligence},
  author={Bariah, Lina and Mongaillard, Thomas and Lasaulce, Samson and Hamidouche, Wassim and Zou, Hang and Debbah, Merouane},
  journal={IEEE Wireless Communications},
  year={2025},
  publisher={IEEE}
}

@article{lewis2020retrieval,
  title={Retrieval-augmented generation for knowledge-intensive nlp tasks},
  author={Lewis, Patrick and Perez, Ethan and Piktus, Aleksandra and Petroni, Fabio and Karpukhin, Vladimir and Goyal, Naman and K{\"u}ttler, Heinrich and Lewis, Mike and Yih, Wen-tau and Rockt{\"a}schel, Tim and others},
  journal={Advances in neural information processing systems},
  volume={33},
  pages={9459--9474},
  year={2020}
}

@inproceedings{brown2020language,
  title     = {Language Models are Few-Shot Learners},
  author    = {Brown, Tom B. and Mann, Benjamin and Ryder, Nick and Subbiah, Melanie and Kaplan, Jared and others},
  booktitle = {Advances in Neural Information Processing Systems (NeurIPS)},
  year      = {2020}
}

@inproceedings{taniuchi2009ieee,
  title     = {{IEEE 802.21}: Media Independent Handover: Features, Applicability, and Realization},
  author    = {Taniuchi, Kenichi and Ohba, Yoshihiro and Fajardo, Victor and Das, Subir and Tauil, Miriam and Cheng, Yuu-Heng and Dutta, Ashutosh and Baker, Donald and Yajnik, Maya and Famolari, David},
  booktitle = {IEEE Communications Magazine},
  year      = {2009}
}

@article{haklay2008openstreetmap,
  title={Openstreetmap: User-generated street maps},
  author={Haklay, Mordechai and Weber, Patrick},
  journal={IEEE Pervasive computing},
  volume={7},
  number={4},
  pages={12--18},
  year={2008},
  publisher={Ieee}
}

@article{mihai2022digital,
  title={Digital twins: A survey on enabling technologies, challenges, trends and future prospects},
  author={Mihai, Stefan and others},
  journal={IEEE Commun. Surveys Tuts.},
  volume={24},
  number={4},
  pages={2255--2291},
  year={2022},
  publisher={IEEE}
}

@article{xu2023digital,
  title={Digital twin-driven collaborative scheduling for heterogeneous task and edge-end resource via multi-agent deep reinforcement learning},
  author={Xu, Chi and others},
  journal={IEEE J. Sel. Areas Commun.},
  volume={41},
  number={10},
  pages={3056--3069},
  year={2023},
  publisher={IEEE}
}

@article{xu2023digital2,
  title={Digital twin and meta rl empowered fast-adaptation of joint user scheduling and task offloading for mobile industrial iot},
  author={Xu, Hansong and Wu, Jun and Pan, Qianqian and Liu, Xing and Verikoukis, Christos},
  journal={IEEE J. Sel. Areas Commun.},
  volume={41},
  number={10},
  pages={3254--3266},
  year={2023},
  publisher={IEEE}
}

@article{jia2023new,
  title={A new virtual network topology-based digital twin for spatial-temporal load-balanced user association in 6G HetNets},
  author={Jia, Pengyi and Wang, Xianbin},
  journal={IEEE J. Sel. Areas Commun.},
  volume={41},
  number={10},
  pages={3080--3094},
  year={2023},
  publisher={IEEE}
}

@article{zhou2023digital,
  title={Digital twin enhanced federated reinforcement learning with lightweight knowledge distillation in mobile networks},
  author={Zhou, Xiaokang and Zheng, Xuzhe and Cui, Xuesong and Shi, Jiashuai and Liang, Wei and Yan, Zheng and Yang, Laurence T and Shimizu, Shohei and Wang, Kevin I-Kai},
  journal={IEEE J. Sel. Areas Commun.},
  volume={41},
  number={10},
  pages={3191--3211},
  year={2023},
  publisher={IEEE}
}

@inproceedings{hoydis2023sionna,
  title={Sionna RT: Differentiable ray tracing for radio propagation modeling},
  author={Hoydis, Jakob and A{\"\i}t Aoudia, Fay{\c{c}}al and Cammerer, Sebastian and Nimier-David, Merlin and Binder, Nikolaus and Marcus, Guillermo and Keller, Alexander},
  booktitle={2023 IEEE Globecom Workshops (GC Wkshps)},
  pages={317--321},
  year={2023},
  organization={IEEE}
}

@inproceedings{khalili2025wcnc,
  author    = {Mohammad Khalili and Marcos Katz and Hazem Sallouha and Sofie Pollin and Konstantin Mikhaylov},
  title     = {Multi-Attribute Handover in Optical Wireless and Radio Frequency Heterogeneous Networks: A Decentralized Approach},
  booktitle = {Proc. IEEE Wireless Communications and Networking Conference (WCNC)},
  pages     = {1--6},
  year      = {2025},
  doi       = {10.1109/WCNC61545.2025.10978689}
}

@article{khalili2025tvt,
  author  = {Mohammad Khalili and Marcos Katz and Konstantin Mikhaylov},
  title   = {Multi-Criteria Handover in Optical Wireless and Radio Frequency Heterogeneous Networks},
  journal = {IEEE Transactions on Vehicular Technology},
  pages   = {1--13},
  year    = {2025},
  doi     = {10.1109/TVT.2025.3621729}
}

@techreport{3gpp_ts_23_501,
  author      = {{3GPP}},
  title       = {{System Architecture for the 5G System (5GS); Stage 2}},
  institution = {{3rd Generation Partnership Project (3GPP)}},
  type        = {{Technical Specification}},
  number      = {{TS 23.501}},
  year        = {2026},
  note        = {{Release 20}}
}

@techreport{3gpp_ts_28_552,
  author      = {{3GPP}},
  title       = {{Management and Orchestration; 5G Performance Measurements}},
  institution = {{3rd Generation Partnership Project (3GPP)}},
  type        = {{Technical Specification}},
  number      = {{TS 28.552}},
  year        = {2026},
  note        = {{Release 20}}
}

@article{Boateng2026LLMCommSurvey,
  title={A Survey on Large Language Models for Communication, Network, and Service Management: Application Insights, Challenges, and Future Directions},
  author={Boateng, Gordon Owusu and others},
  journal={IEEE Commun. Surveys Tuts.},
  volume={28},
  pages={527--566},
  year={2026},
  publisher={IEEE},
  doi={10.1109/COMST.2025.3564333}
}

@article{Zhou2024LLMTelecomSurvey,
  title={Large Language Model (LLM) for Telecommunications: A Comprehensive Survey on Principles, Key Techniques, and Opportunities},
  author={Zhou, Hao and others},
  journal={IEEE Commun. Surveys Tuts.},
  year={2024},
  publisher={IEEE},
  note={Early access},
  doi={10.1109/COMST.2024.3465447}
}

@article{Jiang2024LLMMultiAgentWCM,
  title={Large Language Model Enhanced Multi-Agent Systems for 6G Communications},
  author={Jiang, Feibo and Peng, Yubo and Dong, Li and Wang, Kezhi and Yang, Kun and Pan, Cunhua and Niyato, Dusit and Dobre, Octavia A},
  journal={IEEE Wireless Communications},
  volume={31},
  number={6},
  pages={48--55},
  year={2024},
  publisher={IEEE},
  doi={10.1109/MWC.016.2300600}
}

@article{Zhou2025PromptWCM,
  title={Large Language Models for Wireless Networks: An Overview from the Prompt Engineering Perspective},
  author={Zhou, Hao and Hu, Chengming and Yuan, Dun and Yuan, Ye and Wu, Di and Chen, Xi and Tabassum, Hina and Liu, Xue},
  journal={IEEE Wireless Communications},
  volume={32},
  number={4},
  pages={98--106},
  year={2025},
  publisher={IEEE},
  doi={10.1109/MWC.001.2400384}
}

@article{bilen2022proof,
  author  = {T. Bilen and others},
  title   = {{A Proof of Concept on Digital Twin-Controlled WiFi Core Network Selection for In-Flight Connectivity}},
  journal = {IEEE Commun. Stand. Mag.},
  volume  = {6},
  number  = {3},
  pages   = {60--68},
  month   = sep,
  year    = {2022},
  doi     = {10.1109/MCOMSTD.0001.2100103}
}

@article{cakir2026digital,
  author  = {L. V. Cakir and M. A. Erturk and M. Ozdem and B. Canberk},
  title   = {{Digital Twin-Assisted Handover Scheme for Mobile Networks Using Generative AI}},
  journal = {IEEE Trans. Netw. Service Manag.},
  note    = {early access},
  year    = {2026},
  doi     = {10.1109/TNSM.2026.3690572}
}

@article{tang2025digital,
  author  = {L. Tang and Y. Yi and others},
  title   = {{Digital Twin Construction and Resource Allocation on Internet of Vehicles}},
  journal = {IEEE Internet Things J.},
  volume  = {12},
  number  = {7},
  pages   = {9091--9106},
  month   = apr,
  year    = {2025},
  doi     = {10.1109/JIOT.2024.3507290}
}

@article{guo2022multiattribute,
  title={Multiattribute access selection algorithm for heterogeneous wireless networks based on fuzzy network attribute values},
  author={Guo, Xiaoxue and others},
  journal={IEEE Access},
  volume={10},
  pages={74071--74081},
  year={2022},
  publisher={IEEE}
}

@article{buzzi2016survey,
  author  = {Buzzi, Stefano and others},
  title   = {A Survey of Energy-Efficient Techniques for {5G} Networks and Challenges Ahead},
  journal = {IEEE Journal on Selected Areas in Communications},
  volume  = {34},
  number  = {4},
  pages   = {697--709},
  year    = {2016},
  doi     = {10.1109/JSAC.2016.2550338}
}

@article{roy2025fuzzy,
  title={Fuzzy-{AHP} Based Network Selection in HetNet: An Energy-Efficient and {QoS}-Aware Approach},
  author={Roy, Debabrata and Das, Avirup and Saha, Dibakar},
  journal={IEEE Trans. Green Commun. Netw.},
  year={2025},
  publisher={IEEE}
}

@ARTICLE{zou2025telecomgpt,
  author={Zou, Hang and others},
  journal={IEEE Transactions on Machine Learning in Communications and Networking}, 
  title={{{TelecomGPT}: A Framework to Build Telecom-Specific Large Language Models}}, 
  year={2025},
  volume={3},
  number={},
  pages={948-975},
}

@article{zou2026large,
  title={Large language models in {6G} from standard to on-device networks},
  author={Zou, Hang and others},
  journal={Nature Reviews Electrical Engineering},
  pages={1--12},
  year={2026},
}

@ARTICLE{zou2025access,
  author={{Zou}, Hang and others},
  journal={IEEE Access}, 
  title={GenAINet: Enabling Wireless Collective Intelligence via Knowledge Transfer and Reasoning}, 
  year={2025},
  volume={13},
  number={},
  pages={77764-77777},
}

\vspace{11pt}

\vfill

\end{document}